# The Neutron Star Population in M28: A Joint Chandra/GBT Look at Pulsar Paradise

Eda Vurgun[1], Manuel Linares[1,2], Scott Ransom[3], Alessandro Papitto[4], Slavko Bogdanov[5], Enrico Bozzo[6], Nanda Rea[7,8], Domingo García-Senz[1], Paulo Freire[9], and Ingrid Stairs[10]

[1] Departament de Física, EEBE, Universitat Politècnica de Catalunya, c/Eduard Maristany 16, E-08019 Barcelona, Spain; eda.vurgun@upc.edu
[2] Department of Physics, Norwegian University of Science and Technology, NO-7491 Trondheim, Norway
[3] National Radio Astronomy Observatory, 520 Edgemont Road, Charlottesville, VA 22903, USA
[4] INAF—Osservatorio Astronomico di Roma, Via Frascati 33, I-00040 Monte Porzio Catone, Roma, Italy
[5] Columbia Astrophysics Laboratory, Columbia University, 550 West 120th Street, New York, NY 10027, USA
[6] ISDC Data Centre for Astrophysics, Chemin d'Ecogia 16, CH-1290 Versoix, Switzerland
[7] Institute of Space Sciences (ICE, CSIC), Campus UAB, Carrer de Can Magrans s/n, E-08193, Barcelona, Spain
[8] Institut d'Estudis Espacials de Catalunya (IEEC), Carrer Gran Capita 2–4, E-08034 Barcelona, Spain
[9] Max-Planck-Institut für Radioastronomie, Auf dem Hügel 69, D-53121 Bonn, Germany
[10] Department of Physics and Astronomy, University of British Columbia, 6224 Agricultural Road, Vancouver, BC V6T 1Z1, Canada
Received 2022 May 13; revised 2022 October 26; accepted 2022 October 28; published 2022 December 13

## Abstract

We present the results of a deep study of the neutron star (NS) population in the globular cluster M28 (NGC 6626), using the full 330 ks 2002–2015 ACIS data set from the Chandra X-ray Observatory and coordinated radio observations taken with the Green Bank Telescope (GBT) in 2015. We investigate the X-ray luminosity ($L_X$), spectrum, and orbital modulation of the seven known compact binary millisecond pulsars in the cluster. We report two simultaneous detections of the redback PSR J1824−2452I (M28I) and its X-ray counterpart at $L_X = [8.3 \pm 0.9] \times 10^{31}$ erg s$^{-1}$. We discover a double-peaked X-ray orbital flux modulation in M28I during its pulsar state, centered around pulsar inferior conjunction. We analyze the spectrum of the quiescent NS low-mass X-ray binary to constrain its mass and radius. Using both hydrogen and helium NS atmosphere models, we find an NS radius of $R = 9.2$–$11.5$ km and $R = 13.0$–$17.5$ km, respectively, for an NS mass of $1.4\ M_\odot$ (68% confidence ranges). We also search for long-term variability in the 46 brightest X-ray sources and report the discovery of six new variable low-luminosity X-ray sources in M28.

*Unified Astronomy Thesaurus concepts:* Neutron stars (1108); Millisecond pulsars (1062); Low-mass x-ray binary stars (939)

## 1. Introduction

Neutron stars (NSs) slow down rapidly after birth, reaching spin periods of 0.1–10 s. However, more than 400 millisecond radio pulsars (MSPs) are known in the Galactic field (Manchester et al. 2005; Lorimer 2008, 2019), with spin periods $P_s < 30$ ms. According to the leading theory (Alpar et al. 1982), such fast-spinning NSs are spun up or "recycled" by the accretion of matter in low-mass X-ray binary (LMXB) systems, which are therefore considered the progenitors of MSPs. When the accretion of matter decreases, rotation-powered MSPs can be detected.

Compact binary MSPs are a growing class of pulsars in tight orbits ($P_{\rm orb} \lesssim 1$ day; Roberts 2013) occulted by outflowing plasma during a large fraction of the orbit. Interestingly, the study of this new class of pulsars has revealed two clearly distinct states in quiescence: the disk and pulsar states (Archibald et al. 2009). The disk state is characterized by an intermediate X-ray luminosity ($L_X \sim 10^{33}$ erg s$^{-1}$), strong variability including X-ray mode switching on timescales shorter than the orbital period ($P_{\rm orb}$), and broad, double-peaked (DP) optical emission lines typical of accretion disks (Linares 2014; Linares et al. 2014b). The pulsar state shows radio pulsations and has the lowest X-ray luminosity ($L_X \lesssim 10^{32}$ erg s$^{-1}$). In this state, the companion stars are often strongly irradiated by the relativistic pulsar wind, which has inspired cannibalistic spider nicknames for compact binary MSPs. Black widows (BWs) have very low-mass semidegenerate companions ($<0.1\ M_\odot$) and low X-ray luminosity $L_X \sim 10^{30}$–$10^{31}$ erg s$^{-1}$ (0.5–10 keV). Redbacks (RBs), on the other hand, have more massive nondegenerate companion stars ($0.1$–$0.4\ M_\odot$) and on average, higher $L_X \sim 10^{31}$–$10^{32}$ erg s$^{-1}$.

In the pulsar state the X-ray emission of most RBs and some BWs is predominantly nonthermal, while the intensity (and $L_X$) vary with orbital phase (Bogdanov et al. 2005). This is commonly interpreted as the signature of an intrabinary shock (IBS) between the pulsar and companion winds, with uncertain shape and location, viewed from different angles along the orbit. It has been pointed out that in most cases the orbital-phased light curves show DP maxima centered on the pulsar's inferior conjunction (IC), which suggests that the IBS is curved around the pulsar (Romani & Sanchez 2016; Wadiasingh et al. 2017, 2018; Kandel et al. 2019; van der Merwe et al. 2020).

The two populations of rapidly spinning NSs, MSPs and LMXBs, were definitively connected in spring 2013. A new "transitional MSP" (tMSP) was found in M28 (IGR J18245−2452, or M28I hereafter; Papitto et al. 2013). The system was known as a binary rotation-powered 3.9 ms MSP (Bégin 2006) with an orbital period of 11 hr and then in 2013 April showed a full-fledged accretion outburst with all the characteristics of transient NS LMXBs (i.e., the outburst state with $L_X \sim 10^{34}$–$10^{37}$ erg s$^{-1}$). This provided the strongest evidence in favor of the recycling scenario for MSP formation and raised a flurry of interesting questions about the interaction between the

Table 1
Chandra–ACIS X-Ray Observations of M28 Analyzed in This Work

| Obs. ID | Start Time (MJD) | Date | Exp. Time (ks) | Frame Time (s) | Phase$_{M28I}$[a] | Phase$_{M28H}$[a] |
|---|---|---|---|---|---|---|
| 2684 | 52459.75161 | 2002 Jul 4 | 12.91 | 3.1 | 0.64–0.96 | 0.29–0.45 |
| 2685 | 52490.99057 | 2002 Aug 4 | 13.69 | 3.1 | 0.65–0.98 | 0.93–1.28 |
| 2683 | 52526.70489 | 2002 Sep 9 | 14.3 | 3.1 | 0.44–0.73 | 0.05–0.12 |
| 9132 | 54685.86508 | 2008 Aug 7 | 144.14 | 3.1 | ⋯ | 0.41–1.10 |
| 9133 | 54688.99333 | 2008 Aug 10 | 55.18 | 3.1 | ⋯ | 0.49–1.93 |
| 16748 | 57172.10733 | 2015 May 30 | 30.05 | 3.2 | 0.13–0.86 | 0.21–1.18 |
| 16749 | 57241.84265 | 2015 Aug 7 | 29.93 | 3.2 | 0.92–1.64 | 0.81–1.46 |
| 16750 | 57333.6706057,333.67060 | 2015 Nov 7 | 29.95 | 3.2 | 0.78–1.52 | 0.49–1.53 |

**Note.**
[a] Orbital phase range covered by each observation for M28I (with 10 bins) and M28H (6 bins).

relativistic pulsar wind and its environment. To date, three of the known RBs have been confirmed as transitional MSPs, switching between disk and pulsar states (Bassa et al. 2014; Stappers et al. 2014; Papitto & de Martino 2022).

Galactic globular clusters (GCs) are extremely efficient at forming MSPs and LMXBs due to their high stellar densities (Camilo & Rasio 2005; Verbunt & Lewin 2005). Indeed, more than 200 MSPs are known in GCs (Ransom 2008; Freire 2021). Since many of those MSPs and LMXBs are closely packed within the GC core, their X-ray counterparts can only be fully resolved using Chandra's subarcsecond angular resolution. Messier 28 (NGC 6626, or M28 hereafter) at a distance of 5.5 kpc (Harris 2010) is of particular interest among them, as it hosts one quiescent LMXB (qLMXB hereafter; source 26 from Becker et al. 2003; see also Servillat et al. 2012) and 14 known radio pulsars (7 of which are compact binary MSPs; Freire 2021).

One of the goals of high-energy astrophysics is to determine the mass ($M$) and radius ($R$) of NSs since $M$ and $R$ constrain the equation of state in their interiors. One way of doing so is by fitting the surface thermal X-ray spectra from qLMXBs with NS atmosphere models (Rutledge et al. 2002; Heinke et al. 2006; Steiner et al. 2018). qLMXBs in GCs are good candidates because of their weak magnetic field (B ∼ 10$^{10}$; Gauss Di Salvo & Burderi 2003) and well-known distances to their host clusters (5.5 ± 0.3 kpc for M28; Harris 2010). $M$ and $R$ measurements of NSs in qLMXBs rely on atmosphere modeling. In early studies, it was assumed that the NS atmosphere is composed exclusively of hydrogen since heavier elements are expected to settle quickly below the atmosphere (Rutledge et al. 2002; Heinke et al. 2006). Later work showed that a helium atmosphere gives significant departures in the emergent spectrum and thus systematically affects the inferred $M$ and $R$ (Ho & Heinke 2009; Servillat et al. 2012). In this study, we perform a spectral analysis to constrain $M$ and $R$ for the known qLMXB in M28, using hydrogen and helium atmosphere models and the full available Chandra data set.

In previous Chandra studies of M28, Becker et al. (2003) detected and analyzed 46 relatively bright X-ray sources in detail and found 13 variable sources. Bogdanov et al. (2011) detected and studied 7 of the 12 MSPs known at the time (Bégin 2006) and found indications of orbital variability in the RB MSP PSR J1824−2452H. More recently, Cheng et al. (2020) detected 502 X-ray sources using the full Chandra data set and used them to study the dynamical properties and evolution of M28.

In 2015, we obtained three coordinated Chandra and Green Bank Telescope (GBT) observations of M28 in order to study M28I and the rest of the NS population. Here we report the results of our analysis of the full Chandra–Advanced CCD Imaging Spectrometer (ACIS) data set of the cluster (taken between 2002 and 2015), focusing on the NS population as well as the 46 brightest X-ray sources. We also report several detections of radio-pulsed emission from M28I in our 2015 GBT observations; two of these observations (MJDs 57,172.16 and 57,333.71) are strictly simultaneous with a Chandra X-ray detection. We report the discovery of X-ray orbital modulation in the transitional MSP M28I. We present improved mass and radius constraints from spectral fits of the qLMXB in M28, using both hydrogen and helium NS atmosphere models. We also discover six new variable X-ray sources in the cluster. In Section 2, we describe the observations and data analysis procedure. In Section 3, we present the results of our spectral and temporal analyses of the X-ray sources. In Section 4, we discuss our main results.

## 2. Observations and Data Analysis

### 2.1. Chandra X-Ray Observatory

We analyzed eight observations of M28 collected from the Chandra X-ray Observatory taken between 2002 and 2015 with a total exposure time of 330 ks (see Table 1). We employed the observations performed with the ACIS providing good spectral resolution.[11] We used the CIAO[12] version 4.13 to extract the spectra and light curves (Fruscione et al. 2006).

First, we computed the relative astrometric correction (with the tools *wcs-match* and *wcs-update*) using the longest observation as reference. We created exposure maps in the 0.2−8.0 keV band using *fluximage* and produced point-spread function (PSF) maps using *mkpsfmap* to compute the PSF size at 2.3 keV for a 90% enclosed-counts fraction for each pixel in the image. Then, we created a merged image in the 0.2−8.0 keV band using the *merge-obs* tool (see Figure 1). We also refined the absolute astrometry of the merged X-ray image to compare it with radio pulsar positions. For M28A (for which the radio–X-ray association is well established), we found an offset between its radio position and its X-ray counterpart of $\Delta_{R.A.} = 0.''1$ and $\Delta_{decl.} = -0.''3$ in R.A. and decl., respectively. We then applied this correction to the astrometric frame of the X-ray data set. We took the centroid coordinates of the X-ray counterparts of I, L, and their positional uncertainties from Becker et al. (2003). Shifted X-ray positions of C, D, E, G, M, and their positional uncertainties are taken from Cheng et al. (2020). The rest of the X-ray positions and their positional uncertainties are obtained in this work. We estimated positional uncertainties (95% confidence) using an empirical relation from Hong et al. (2005). Then, we calculated the angular separation between the radio and the X-ray coordinates (all reported in Table A1).

---
[11] ACIS Instrument Information: https://cxc.harvard.edu/proposer/POG/html/chap6.html.
[12] Chandra Interactive Analysis of Observations, available at https://cxc.harvard.edu/ciao/.

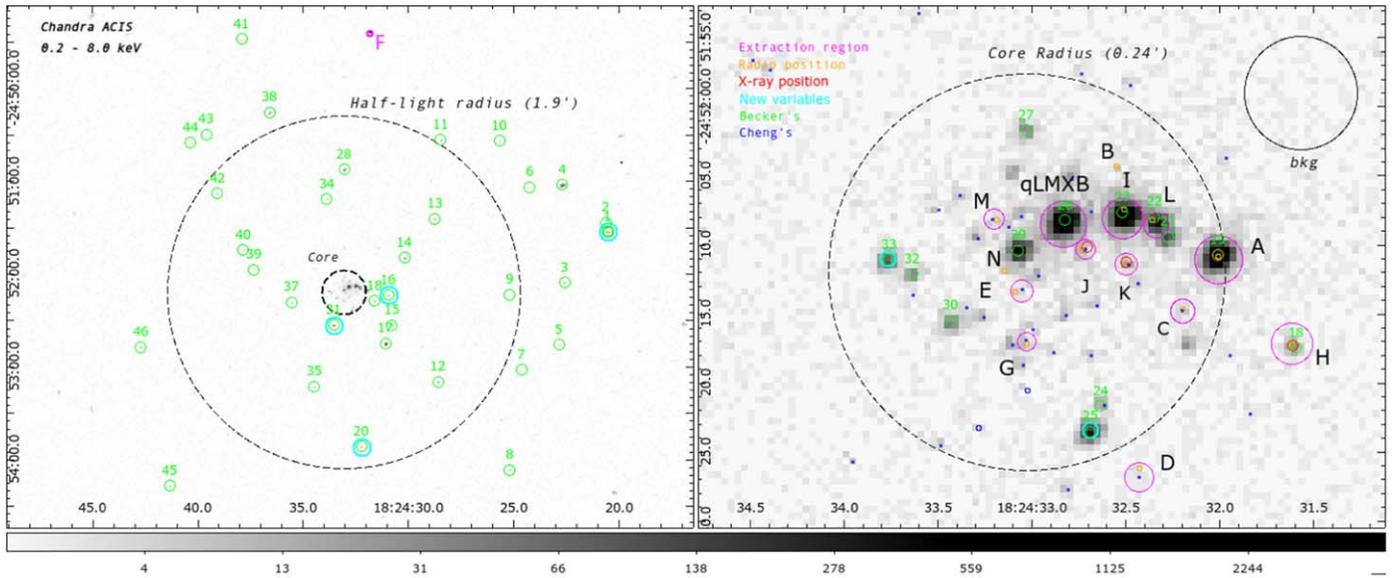

**Figure 1.** Full-band (0.2–8.0 keV) Chandra merged image of the GC M28. Left panel: The black dashed circle shows the half-light radius of 1′.9 of the cluster (J1824–2452F is the only known pulsar outside this circle). The green circles show the 46 X-ray sources detected by Becker et al. (2003). Right panel: the black dashed circle shows the core of M28 with a 0′.24 radius (Harris 1996). The small blue circles show the X-ray sources detected by Cheng et al. (2020). Magenta circles show the extraction regions, and cyan circles show the new variable sources detected in this work. Red circles show the X-ray positions obtained in this work. Orange circles show the exact radio positions of the known radio pulsars. M's and N's radio positions are taken from Douglas et al. (2022).

Following the standard CIAO data analysis threads,[13] we extracted each spectrum using the *specextract* tool. Response matrices and ancillary response files were generated using the *mkacisrmf* and *mkarf* tools, respectively. We created circular regions for the source extraction with radii between 0.''7 and 1.''7, depending on the nearby surroundings. We used a region of 4″ radius for the background extraction from a source-free part of the same chip. We fitted our spectra in the 0.2−8.0 keV energy range using XSPEC 12.11.0 (Arnaud 1996), generally employing a phenomenological-absorbed power-law model. For the brightest pulsar M28A, we included the Chandra pileup model in the spectral fitting (Davis 2001; Heinke et al. 2006; Ho & Heinke 2009; Suleimanov et al. 2014). For M28I in the disk state, we verified that our reported spectral parameters are not affected by pileups. We fit the spectra of the bright sources, grouping them to a minimum of 15 counts per channel and using chi-squared statistics. We fit the spectra of the faint sources (less than 150 net counts), grouping them so that each channel contains a minimum of one count and using Cash's C-statistic. We used the *tbabs* model to account for interstellar absorption (Wilms et al. 2000). Individual observations were fitted simultaneously, leaving the spectral parameters free to vary when possible. For the power-law fits, we kept constant the equivalent hydrogen column density ($N_H$) at $0.25 \times 10^{22}$ cm$^{-2}$, thus assuming that it did not vary between 2002 and 2015. All spectral fits and luminosities reported herein use a distance of 5.5 kpc (Harris 1996). For the faint sources, we employed average-fit results for each source and set upper limits on the $L_X$ for the observations where these are not detected or had very low counts. We calculated the net counts, taking the region size of 1.''5 for each source (and 3.''1 for M28F since it is outside the center regarding PSF size) and placing 90% confidence level (c.l.) in the range of 0.5–10.0 keV. Then, we divided the upper limits of the net counts by the exposure time of every single Chandra observation. Then, we calculated fluxes using WebPIMMS[14] giving $N_H$ and $\Gamma$ parameters from the average-fit results.

In order to constrain the NS mass ($M$) and radius ($R$), we fitted the spectrum of the qLMXB using hydrogen (NSATMOS) and helium (NSX) NS atmosphere models, including the Chandra pileup model (Davis 2001; Heinke et al. 2006; Ho & Heinke 2009; Suleimanov et al. 2014). After verifying that they are consistent within the errors, we tied all parameters between different data sets and fixed the normalization to 1, thereby assuming that all the NS surface is emitting. We also kept the frame time frozen at 3.14 s for the observations between 2002 and 2008 and 3.24 s for the 2015 observations.

For the orbital phase-folded light curves, we studied the five brightest compact binary MSPs, namely M28G, M28H, M28I, M28J, and M28L (two RBs and three BWs; see Table 2). We applied barycentric corrections to the photon arrival times in each event and aspect file using the *axbary* tool, and we computed the orbital phase using *dmtcalc*. We computed the phases using the $P_{orb}$ and the epoch of zero mean anomaly ($T_0$)

---

[13] Chandra Interactive Analysis of Observations, available at https://cxc.cfa.harvard.edu/ciao/threads/pointlike/.

[14] Portable, Interactive Multi-Mission Simulator https://heasarc.gsfc.nasa.gov/cgi-bin/Tools/w3pimms/w3pimms.pl.

Table 2
Counts, Rates, and Orbital Parameters for the Known NS Systems in M28

| Source | Net Count[a] | Count Rate (Count s$^{-1}$) | log $P_B$[S/N][b] | Type[c] | $T_0$[d] (MJD) | $P_s$ (ms) | $P_{orb}$ (hr) | $M_c$[e] ($M_\odot$) | Reference |
|---|---|---|---|---|---|---|---|---|---|
| q | 13418 | 4e-02 | −[36] | qLMXB | ⋯ | ⋯ | ⋯ | ⋯ | 1,2 |
| A | 9031 | 3e-02 | −[30] | MSP | i | 3.05 | i | i | 3,8 |
| B | 20 | 6e-05 | ⋯ | MSP | i | 6.55 | i | i | 3,5 |
| C | 27 | 9e-05 | <−5 | MSP, eccentric | ⋯ | 4.16 | 193.8674 | 0.30 | 3,5 |
| D | 18 | 5e-05 | <−5 | young, eccentric | ⋯ | 79.83 | 729.8762 | 0.45 | 3,5 |
| E | 20 | 7e-05 | <−5 | MSP | i | 5.42 | i | i | 3,5 |
| F | 28 | 9e-05 | <−5 | MSP | i | 2.45 | i | i | 3,5 |
| G | 23 | 7e-05 | <−5 | BW | 53629.071809(10) | 5.91 | 2.51000803(17) | 0.01 | 3,5 |
| H | 117 | 4e-04 | −[3] | RB | 53755.2263988(13) | 4.63 | 10.4406611(7) | 0.20 | 3,5 |
| I | 10301 | 3e-02 | −[7] | RB | 56395.216893(1) | 3.93 | 11.025781(2) | 0.20 | 3,4,5 |
| J | 55 | 2e-04 | <−5 | BW | 53832.2815822(36) | 4.04 | 2.33835171(9) | 0.01 | 3,5 |
| K | 52 | 2e-04 | <−5 | MSP | ⋯ | 4.46 | 93.8482 | 0.16 | 3,6 |
| L | 1347 | 4e-03 | −[5] | BW | ⋯ | 4.10 | 5.4170 | 0.02 | 3,6 |
| M | 27 | 8e-05 | <−5 | BW | 56451.272704(15) | 4.78 | 5.82046126(3) | 0.011[f] | 3,7 |
| N | ⋯ | ⋯ | ⋯ | BW | 56451.2896713(15) | 3.35 | 4.76383956(3) | 0.019[f] | 7 |

**Notes.**
[a] Background-subtracted net counts extracted in the 0.2−8.0 keV band.
[b] Logarithm of the binomial no-source probability, taken from Cheng et al. (2020). Sources q,A,H,I,L are indicated by S/N ratio from Becker et al. (2003).
[c] Types are indicated as follows: RB =redback; BW =black widow; MSP =millisecond pulsar.
[d] Epoch of zero mean anomaly.
[e] Companion mass calculated assuming a pulsar mass of 1.35 $M_\odot$ and an inclination of 60°.
[f] The minimum companion mass calculated assuming a pulsar mass of 1.4 $M_\odot$; i = isolated pulsar.
**References.** (1) Becker et al. (2003), (2) Servillat et al. (2012), (3) Freire (2021), (4) Papitto et al. (2013), (5) Bégin (2006), (6) Bogdanov et al. (2011), (7) Douglas et al. (2022), (8) Lyne et al. (1987).

of each MSP as measured from radio-timing observations (see Table 2). Thus, we define $T_0$ and orbital phase zero as the epoch of the ascending node of the pulsar. Finally, we extracted the phase-binned light curves using *dmextract*. In order to obtain the correct count rates, we calculated effective exposure times from the good time intervals for each phase bin.

### 2.2. Green Bank Telescope

As part of a long-term GC pulsar monitoring program with the Green Bank Telescope (GBT), we have observed or acquired archival data for M28 on nearly 100 occasions between 2005 and the present day (PI: Ransom). The vast majority of those observations were centered near either 1.5 GHz (i.e., *L* band) or 2.0 GHz (i.e., *S* band) with a small number of observations using the 820 MHz receiver. Before 2010, the observations used the GBT Spigot (Kaplan et al. 2005) and up through mid-2021 used GUPPI (DuPlain et al. 2008) in a coherently dedispersed high time-resolution (i.e., 10.24 $\mu$s with 512 frequency channels) search mode. For this paper we are primarily concerned with the more recent GUPPI observations, which we partially integrated both in time (by a factor of 4) and frequency (also by a factor of 4, dedispersing the channels incoherently) to give us 40.96 $\mu$s total-intensity samples in 128 frequency channels covering 800 MHz of radio bandwidth, of which ∼650 MHz was typically usable due to radio frequency interference. More information is available about the archival data and its processing in Douglas et al. (2022). We present the radio positions of the 14 known pulsars in M28 in Table A1. Full GBT timing results will be reported elsewhere.

In 2015, we obtained three GBT+GUPPI observations at 2.0 GHz (*S-band*) of about an 8 hr duration (program ID GBT14B−453) that were specifically coordinated with Chandra. Those observations were processed in the same manner as the archival observations, although a more sophisticated method of searching for M28I was used for these data. Two of these GBT observations, starting on MJDs 57172.145822 and 57333.69797557,333.697975 (2015 May 30 and 2015 November 7) were strictly simultaneous with Chandra observations.

Most of the archival data in the past had been searched for detections of M28I by blindly searching the dedispersed time series at the known dispersion measure (DM) of M28I of ∼119 pc cm$^{-3}$ using standard Fourier-domain techniques with PRESTO (Ransom 2011). Alternatively, we used detections on nearby days to determine the instantaneous orbital phase (i.e., time of the ascending node $T_0$) via pulsar timing and then folded the data since as a transitional RB pulsar, there is significant variation in the orbital period (and therefore phase) with time. For the data tied more closely with this project, we used the package SPIDER_TWISTER,[15] which brute-force folds the radio data over a range of $T_0$ values and reports the most significant detection along with the best $T_0$ value. M28I was detected multiple times as a radio pulsar in 2015 using SPIDER_TWISTER, including on both of the days with simultaneous Chandra observations.

### 3. Results

We detected 12 (A, C, D, E, F, G, H, I, J, K, L, and M) of the 14 known pulsars in the cluster by cross-correlating the significant source detections reported by Cheng et al. (2020) with the radio-timing positions of these pulsars (see Figure 1). Indeed, for all pulsars except B their excess counts have a probability <10$^{-5}$ of being produced by background fluctuations (Cheng et al. 2020). We did not detect the newly discovered pulsar N, but we were able to measure $L_X$ for the new MSP M28M (Douglas et al. 2022). In Table 2, we give the wide range of net counts (18–13418) and count rates ($6 \times 10^{-5}$–$4 \times 10^{-2}$ c s$^{-1}$) for the NSs studied in this work.

---
[15] http://alex88ridolfi.altervista.org/pagine/pulsar_software_SPIDER_TWISTER.html

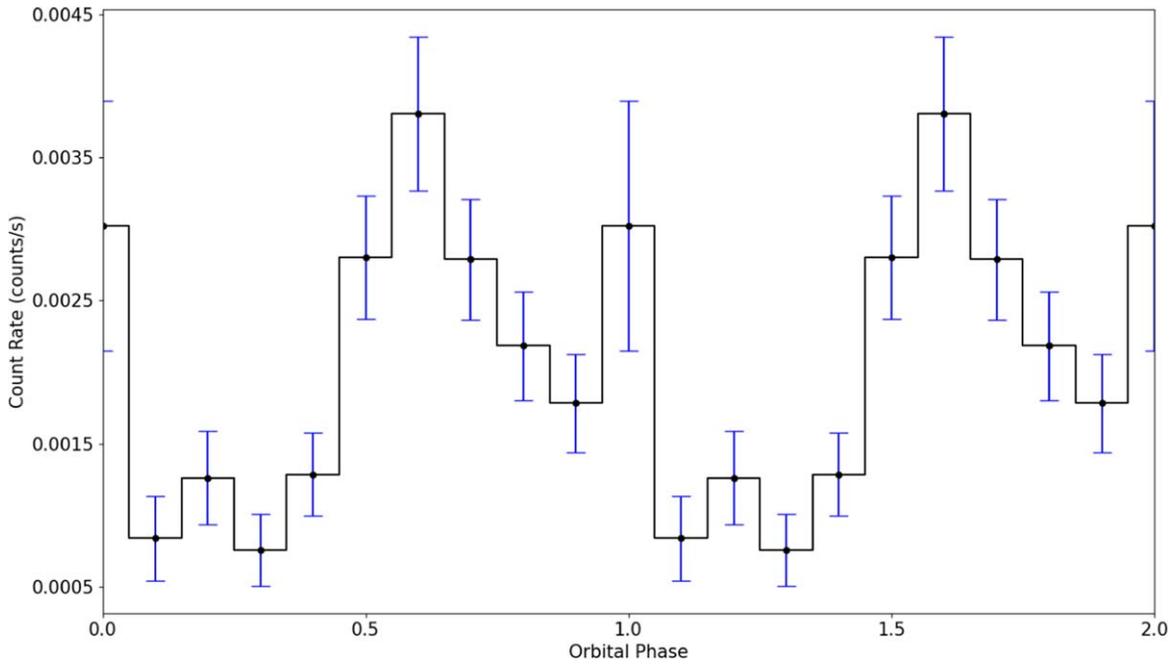

**Figure 2.** Orbital X-ray light curve of M28I in the 0.2–8.0 keV band including the observations taken in 2002 and 2015 when this transitional (and RB) MSP was in the pulsar state (Section 3.1.1). Two cycles are shown for clarity.

We give the radio and X-ray source positions with their uncertainties in Table A1 of the Appendix. In all 12 detected X-ray counterparts to the known pulsars in M28, the X-ray positions agree (within $2\sigma$) with the much more precise radio locations. We quantified the uncertainty in the radio-X-ray cross correlation following Bogdanov et al. (2011); we applied multiple random offsets to the radio pulsar positions (of 2″.5–5″ in R.A. and decl.) and compared these with the X-ray image (Figure 1). We find only one source match per offset due to chance out of the 11 known pulsars in the core of M28 (where match is defined as coordinate agreement within $2\sigma$).

### 3.1. X-Ray Orbital Variability of Spiders

#### 3.1.1. The Transitional MSP M28I

We extracted source counts from the transitional MSP M28I in the 0.2–8.0 keV range (without background subtraction), including the 2002 and 2015 observations when the source was in the pulsar state (for a total exposure of 131 ks, i.e., 3.3 times $P_{orb}$). We note that this accumulated exposure time in the pulsar state has increased by 220% with respect to 2002, thanks to our 2015 observations. Furthermore, the three observations taken in 2015 cover altogether the full orbital phase range, as can be seen in Table 1.

We discover and report X-ray orbital modulation of M28I during the pulsar state, shown in Figure 2. We find evidence for a DP light curve with two maxima at orbital phases 0 and 0.6 and a broad minimum around phase 0.25. We extracted a light curve with eight bins per orbit and found that this double peak is still apparent. This is consistent with the orbital modulation of most RBs: a DP maximum centered around the IC of the pulsar and a minimum at the pulsar's superior conjunction (SC; Wadiasingh et al. 2017). The peak-to-peak semiamplitude of the modulation is 0.0015 c s$^{-1}$, i.e., about 71% of the average count rate (0.0022 c s$^{-1}$) after subtracting the background rate ($9.1 \times 10^{-5}$ c s$^{-1}$). The fractional semiamplitude for some other RBs is typically around 50% in previous studies (Bogdanov et al. 2011; Hui et al. 2015).

#### 3.1.2. The Redback MSP M28H

We extracted the counts from the RB MSP M28H in the 0.2 −10.0 keV range, including all observations with a total exposure of 330 ks. The X-ray orbital variability of the RB MSP M28H was studied by Bogdanov et al. (2011) using the 2002 and 2008 observations. They found a minimum around orbital phase 0.25 and a maximum at phase 1.0. We performed the same analysis as explained in Section 2 in order to compare with their results, using six orbital phase bins.

Our results are shown in Figure 3: we find a broad minimum in the orbital X-ray light curve at phase 0–0.4 and a maximum around phase 0.75 (pulsar at IC). The peak-to-peak semiamplitude of the modulation is 0.0004 c s$^{-1}$, corresponding to a fractional amplitude of 80% of the average count rate (0.0006 c s$^{-1}$) after correcting the background rate (0.0001 c s$^{-1}$). The light-curve shape is consistent with the results of Bogdanov et al. (2011), with a difference in the peak phase of about 0.25. In this case we do not find a DP light curve within the lower-phase resolution imposed by the lower X-ray luminosity of M28H compared to M28I in the pulsar state.

#### 3.1.3. The Black Widow MSPs M28G, M28J, and M28L

We performed the same orbital-phased light-curve analysis for the BW MSPs M28G and M28J, including all observations (Figure 4). These are fainter and less luminous than the RBs (with typical $L_X < 10^{31}$ erg s$^{-1}$) so we used in this case four orbital phase bins.

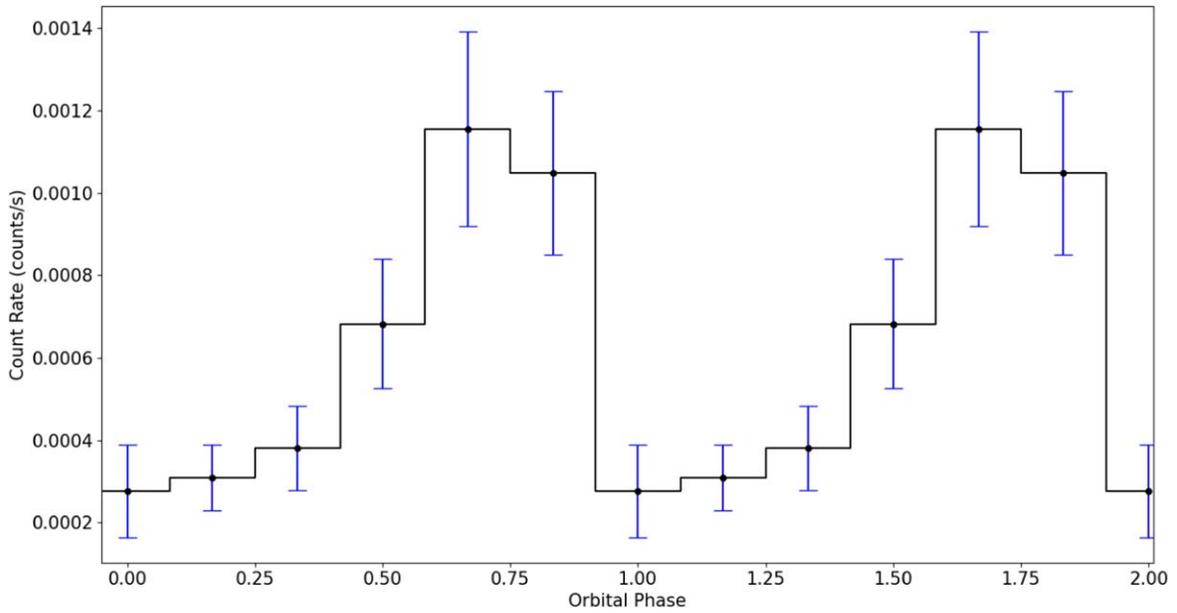

**Figure 3.** Orbital X-ray light curve of M28H in the 0.2–10.0 keV band including all observations. Two cycles are shown for clarity.

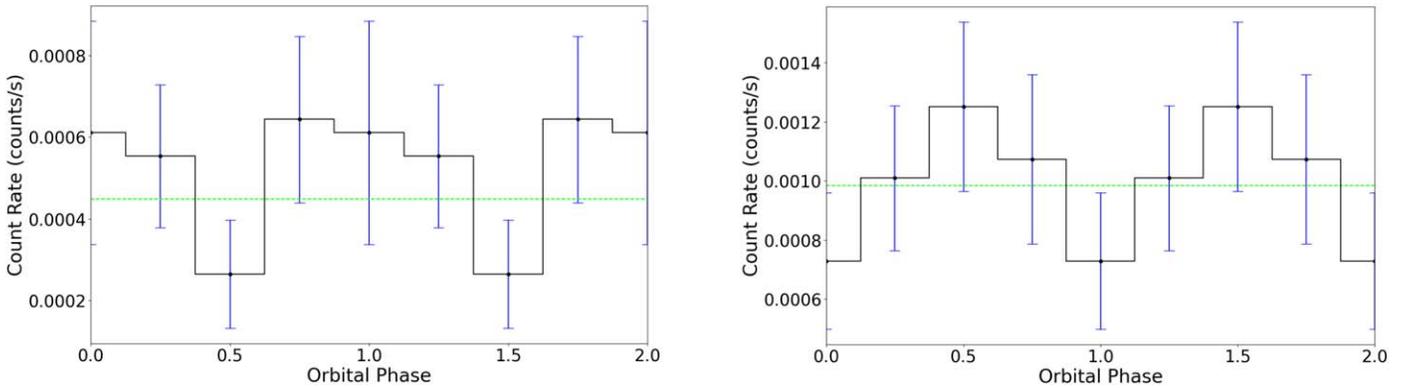

**Figure 4.** Left panel: Chandra–ACIS-S merged two orbital-cycle light curves of M28G with bin number four including all observations. Right panel: Chandra–ACIS-S merged two orbital-cycle light curves of M28J with bin number four including all observations.

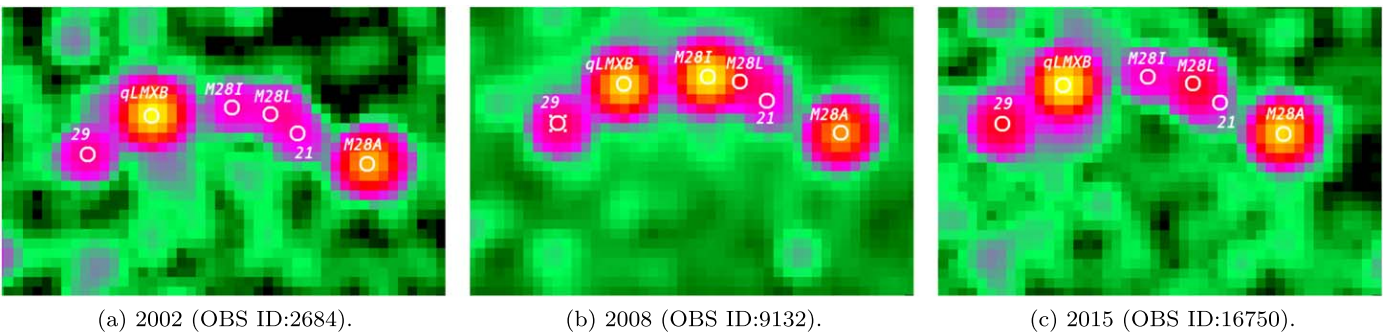

(a) 2002 (OBS ID:2684).  (b) 2008 (OBS ID:9132).  (c) 2015 (OBS ID:16750).

**Figure 5.** Chandra–ACIS archival observations of the core of M28 in three epochs taken in 2002, 2008, and 2015 from left to right. White circles here in the core show the sources from Becker et al. (2003) from the right to the left: M28A, S21, M28L, M28I, qLMXB, and S29. We see a blend of three sources: M28I, M28L, and S21.

The count rates of these two BWs in the four phase bins are approximately constant, as can be seen in Figure 4. We fit both X-ray orbital light curves with a constant and find a $\chi^2/dof$ of 3.5/3 and 2.2/3 for M28G and M28J, respectively. We conclude that these two BWs show no evidence of X-ray orbital variability within the currently available data.

We did not investigate X-ray orbital modulation for the BW MSP M28L since it is suffering from the contamination of the very nearby RB M28I. This effect is even clearer in the 2008 epoch when M28I was in the disk state, as seen in Figure 5. The radio position of M28L (see Table A1) suggests that its count rates are strongly exposed to source confusion. Indeed, the distance between the radio positions of these two sources is $1\rlap{.}''8$, while the PSF size (FWHM) is $1\rlap{.}''5$ assuming a circular Gaussian representing the PSF.[16]

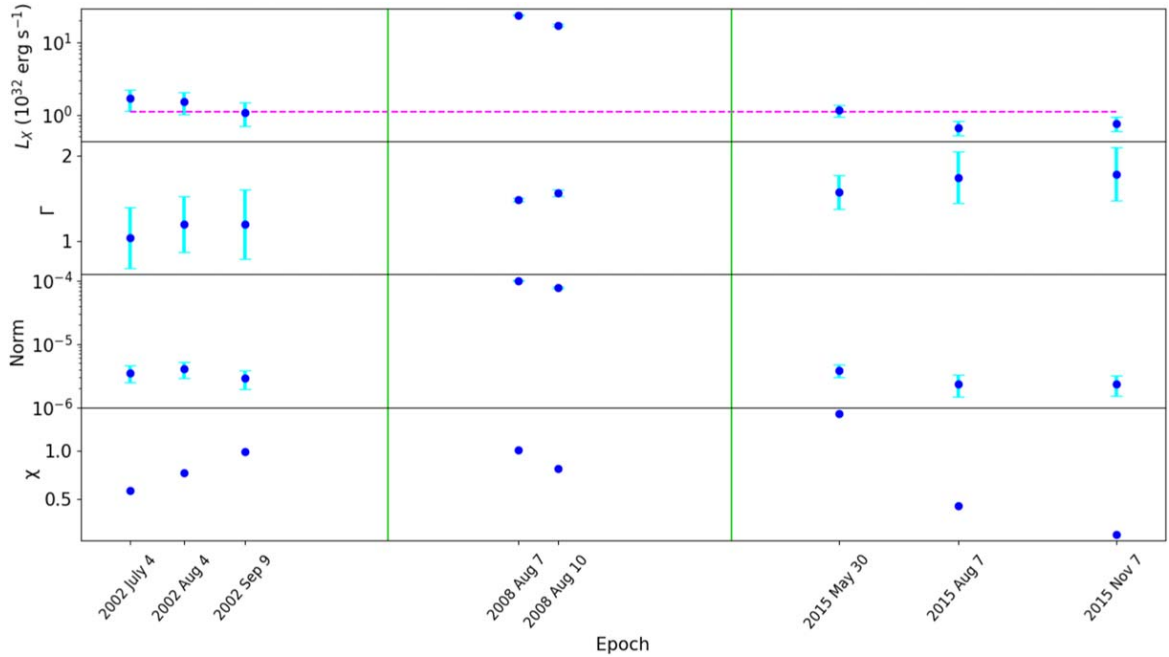

**Figure 6.** Variability of M28I. Top panel: luminosity in $10^{32}$ erg s$^{-1}$. Pink dashed line shows the average luminosity of $1.11 \times 10^{32}$ erg s$^{-1}$. Second panel: photon index parameter of the power-law model. Third panel: normalization parameter of the power-law model. Bottom panel: reduced chi-squared. The green vertical lines separate the epochs visually.

### 3.2. Spectra and Luminosities

We analyzed the spectra of the spider MSPs and the other known pulsars in M28, respectively, for each observation where they are detected. We fit all the spectra from each source jointly within XSPEC, keeping the power-law index and normalization linked between different observations and the $N_H$ frozen to the cluster value. We thereby find the 2002–2015 average flux, $L_X$, and photon index for each system, which we present in Table 3. We find photon indices in the range 1–4 and $L_X$ between $9 \times 10^{29}$ erg s$^{-1}$ and $2 \times 10^{33}$ erg s$^{-1}$ for the full variety of pulsar types and states (see the Appendix for the spectra). Next, we present our results for the transitional MSP and the other known compact binary MSPs in more detail.

We inspected the spectra of the transitional and RB MSP M28I by dividing them into three epochs: 2002, 2008, and 2015 (Table 3). We find that M28I was back in the pulsar state in 2015 with $L_X = [8.3 \pm 0.9] \times 10^{31}$ erg s$^{-1}$, an X-ray luminosity similar to that measured in the 2002 observations, and a photon index $\Gamma = 1.7 \pm 0.2$. This is the lowest $L_X$ measured for M28I to date (see Figure 6). From our reanalysis of the 2002 observations, we measure $L_X = [1.3 \pm 0.2] \times 10^{32}$ erg s$^{-1}$ (slightly lower than that measured by Linares et al. (2014a) yet consistent at the $2\sigma$ confidence level). We measure $L_X = [2.14 \pm 0.04] \times 10^{33}$ erg s$^{-1}$ from the 2008 spectra of M28I, i.e., a disk-state luminosity consistent with the findings of Linares et al. (2014a). We also analyzed the spectra from 2002 and 2015 jointly and find an average pulsar-state luminosity of $[1.1 \pm 0.2] \times 10^{32}$ erg s$^{-1}$ and a photon index of $1.3 \pm 0.3$. We inspected the hardness ratio, taking the energy bands 0.2–2.0 keV and 2.0–8.0 keV and found no significant spectral variability along the orbit.

During the X-ray association analysis, we find that the radio position of M28L agrees with source 22 from Becker et al. (2003; see Table A1 in the Appendix). Thus, we study the possible X-ray counterpart of the BW M28L using a $0\rlap{.}''9$ radius region, which includes the radio position and source 22 from Becker et al. (2003; manually centered to minimize contamination from nearby sources). We find an average luminosity of $L_X = [1.8 \pm 0.1] \times 10^{32}$ erg s$^{-1}$ and a photon index of $1.55 \pm 0.06$ (see Table 3 and Figure A4). As mentioned above and as noted previously (e.g., Bogdanov et al. 2011), this region is severely crowded, and there may be contamination from other sources (mainly M28I and perhaps also source 21; see Figure 5). Thus, this luminosity must be interpreted with care, and we consider the counterpart to M28L as tentative.

The spectra of the RB M28H are well fitted by the power-law model yielding an $L_X = [2.3 \pm 0.4] \times 10^{31}$ erg s$^{-1}$ with a photon index $\Gamma = 1.0 \pm 0.2$. We set an upper limit on $L_X$ for the second 2015 observation, which appears consistent with a constant luminosity (see Figure A4). For the BWs M28G and M28J, we measure average luminosities of $L_X = [1.7 \pm 0.6] \times 10^{30}$ erg s$^{-1}$ and $[5.2 \pm 1.0] \times 10^{30}$ erg s$^{-1}$, respectively. We find photon indices in the 2.5–4 range (Table 3), indicating softer spectra than the redbacks above. We analyzed the spectrum of the newly discovered BW M28M using the best available data from the 2008 fitting with a power-law model (Figure A3). The best fit yields $L_X = [2.7 \pm 0.7] \times 10^{30}$ erg s$^{-1}$, which is similar to the measured luminosities of the other BWs in M28 and in the GC 47 Tucanae (Bogdanov et al. 2006). We also set upper limits on $L_X$ for the observations where these are not detected (see Figure A4).

---
[16] https://cxc.cfa.harvard.edu/ciao/ahelp/psfsize_srcs.html.

Table 3
Results of the Averaged Spectral Fits for the Known Pulsars Detected by Chandra

| Name | $\Gamma$ | Flux[a] (erg cm$^{-2}$ s$^{-1}$) | $L_X$ (erg s$^{-1}$) | Fit Statistic/dof[e] |
|---|---|---|---|---|
| M28A | $1.33 \pm 0.03$ | $[4.6 \pm 0.1] \times 10^{-13}$ | $[1.66 \pm 0.06] \times 10^{33}$ | $446/480(\chi^2)$ |
| M28C | $2.8 \pm 0.9$ | $[6.0 \pm 2.0] \times 10^{-16}$ | $[2.0 \pm 0.7] \times 10^{30}$ | $11/16(CS)$ |
| M28D | $4.2 \pm 1.3$ | $[3.0 \pm 1.0] \times 10^{-16}$ | $[9.3 \pm 4.3] \times 10^{29}$ | $14/11(CS)$ |
| M28E | $2.6 \pm 0.8$ | $[5.0 \pm 2.0] \times 10^{-16}$ | $[1.8 \pm 0.7] \times 10^{30}$ | $19/18(CS)$ |
| M28F | $2.55 \pm 0.9$ | $[3.0 \pm 3.0] \times 10^{-16}$ | $[1.3 \pm 1.2] \times 10^{30}$ | $20/26(CS)$ |
| M28G | $3.5 \pm 0.7$ | $[5.0 \pm 2.0] \times 10^{-16}$ | $[1.7 \pm 0.6] \times 10^{30}$ | $23/18(CS)$ |
| M28H | $1.0 \pm 0.2$ | $[6.0 \pm 1.0] \times 10^{-15}$ | $[2.3 \pm 0.4] \times 10^{31}$ | $99/101(CS)$ |
| M28I-p[b] | $1.1 \pm 0.2$ | $[4.0 \pm 0.8] \times 10^{-14}$ | $[1.44 \pm 0.3] \times 10^{32}$ | $67/82(CS)$ |
| M28I-d[c] | $1.51 \pm 0.02$ | $[5.9 \pm 0.1] \times 10^{-13}$ | $[2.14 \pm 0.04] \times 10^{33}$ | $485/367(\chi^2)$ |
| M28I-p[d] | $1.7 \pm 0.2$ | $[2.3 \pm 0.3] \times 10^{-14}$ | $[8.3 \pm 1.0] \times 10^{31}$ | $16/12(\chi^2)$ |
| M28J | $2.8 \pm 0.3$ | $[1.0 \pm 0.3] \times 10^{-15}$ | $[3.6 \pm 1.0] \times 10^{30}$ | $11/13(CS)$ |
| M28K | $2.9 \pm 0.4$ | $[1.1 \pm 0.2] \times 10^{-15}$ | $[4.1 \pm 0.8] \times 10^{30}$ | $35/36(CS)$ |
| M28L | $1.55 \pm 0.06$ | $[5.0 \pm 0.8] \times 10^{-14}$ | $[1.8 \pm 0.3] \times 10^{32}$ | $67/71(\chi^2)$ |
| M28M | $3.6 \pm 1.3$ | $[7.4 \pm 1.9] \times 10^{-16}$ | $[2.7 \pm 0.7] \times 10^{30}$ | $4/12(CS)$ |

**Notes.**
[a] Unabsorbed flux in the 0.5–10.0 keV band.
[b] Includes 2002 observations (M28I-p: pulsar state).
[c] Includes 2008 observations (M28I-d: disk state).
[d] Includes 2015 observations (M28I-p: pulsar state).
[e] CS and $\chi 2$ indicate the fit obtained with C-statistic and chi-squared statistic, respectively.

### 3.3. Pulsed Radio Emission from M28I

As discussed in Section 2.2, we reanalyzed seven different GBT observations of M28 from 2015, at MJDs 57137.49, 57172.16, 57186.14, 57187.13, 57261.90, 57333.71, and 57382.80, spanning from 2015 April 25 to 2015 December 26 using SPIDER_TWISTER and PRESTO (see Figure 7). M28I was detected in each of the observations, although with large amounts (i.e., factors of several) of flux variability, even when the highly irregular eclipses had not completely eliminated the pulsed radio emission in portions of the scans. Since two of these observations (57,172.16 and 57,333.71) were simultaneous with those from Chandra, the pulsar was definitively in the active radio pulsar state at the time of the X-ray observations and likely throughout most of 2015. The radio-timing observations of M28 and detections of M28I from our GBT campaign are shown in Figure 12.

### 3.4. Thermal X-Rays from the qLMXB: Mass and Radius Constraints

We present the X-ray spectral analysis of the qLMXB in M28 and the resulting NS mass ($M$) and radius ($R$) constraints. Measuring the X-ray flux of the qLMXB, we found that the source shows no significant variability across observations between 2002 and 2015. We find a 25% decrease in count rate in the 2015 observations, which we attribute to molecular[17] contamination of the ACIS detector (see Figure A5). We analyze the full Chandra data set for a total exposure time of 330 ks (39% longer than what was available for previous studies; Servillat et al. 2012). The increase in collected net counts is lower (30%) due to the drop in count rate mentioned above.

In order to study the effect of different atmosphere compositions on $M$–$R$ constraints, we performed the spectral fits using two different models, NSATMOS (Ho & Heinke 2009) and NSX (Heinke et al. 2006), which model a hydrogen and helium NS atmosphere, respectively. These models are valid for negligible magnetic fields (less than $10^9$ G) in agreement with the weak fields expected for NSs in qLMXBs (Di Salvo & Burderi 2003). We also included the pileup component in every spectral fit (see Section 2 for details on the spectral-fitting procedure). We show the folded X-ray spectra, best-fit models, and residuals in Figure 8.

We first fitted the 2002–2008 spectra in order to compare directly with Servillat et al. (2012). We left the $\alpha$ parameter of the pileup model as well as $N_H$ free to vary, in order to include their uncertainties in our results for $M$ and $R$. For the hydrogen and helium atmosphere models, our results are consistent with the $M$ and $R$ constraints of Servillat et al. (2012) within the errors (see panels (a) and (c) in Figure 9).

From now on, we present the results of our best fits to the full data set. The 68%, 90%, and 99% confidence regions in the $M$–$R$ plane are shown in Figure 9 for the hydrogen and helium model fits. Next we present single parameter constraints at the 68% confidence level. From our spectral fits with the hydrogen atmosphere model, we find that $R$ is between 9.2 and 11.5 km for the once canonical NS mass of 1.4 $M_\odot$. We found $N_H = [0.32 \pm 0.02] \times 10^{22}$ cm$^{-2}$ and the temperature $T = 0.13 \pm 0.01$ keV. For the helium atmosphere model, which is performed with the same fitting procedure as hydrogen, we found higher radii $R = 13.0$–17.5 km for $M = 1.4$ M$_\odot$ at the same confidence level with $N_H = [0.35 \pm 0.02] \times 10^{22}$ cm$^{-2}$ and $T = 0.10 \pm 0.01$ keV. The 0.5–10.0 keV absorbed flux of the source (after removing pileup effect) is $[1.8 \pm 0.2] \times 10^{-13}$ erg s$^{-1}$ cm$^{-2}$, which corresponds to $L_x = [6.5 \pm 0.7] \times 10^{32}$ erg s$^{-1}$.

### 3.5. Search for Long-term Variability

We searched for a long-term $L_x$ variability on timescales of years (2002–2008–2015) using the 46 X-ray sources detected by Becker et al. (2003). We performed a spectral analysis of all 46 X-ray sources fitting the spectrum of each observation with a simple power-law model. Then, we estimated the significance (*S*) of the flux variations as

$$S = \frac{F_{\max} - F_{\min}}{(EF_{\max}^2 + EF_{\min}^2)^{1/2}}. \tag{1}$$

---
[17] https://cxc.harvard.edu/proposer/POG/html/chap6.html#tth_sEc6.5.1.

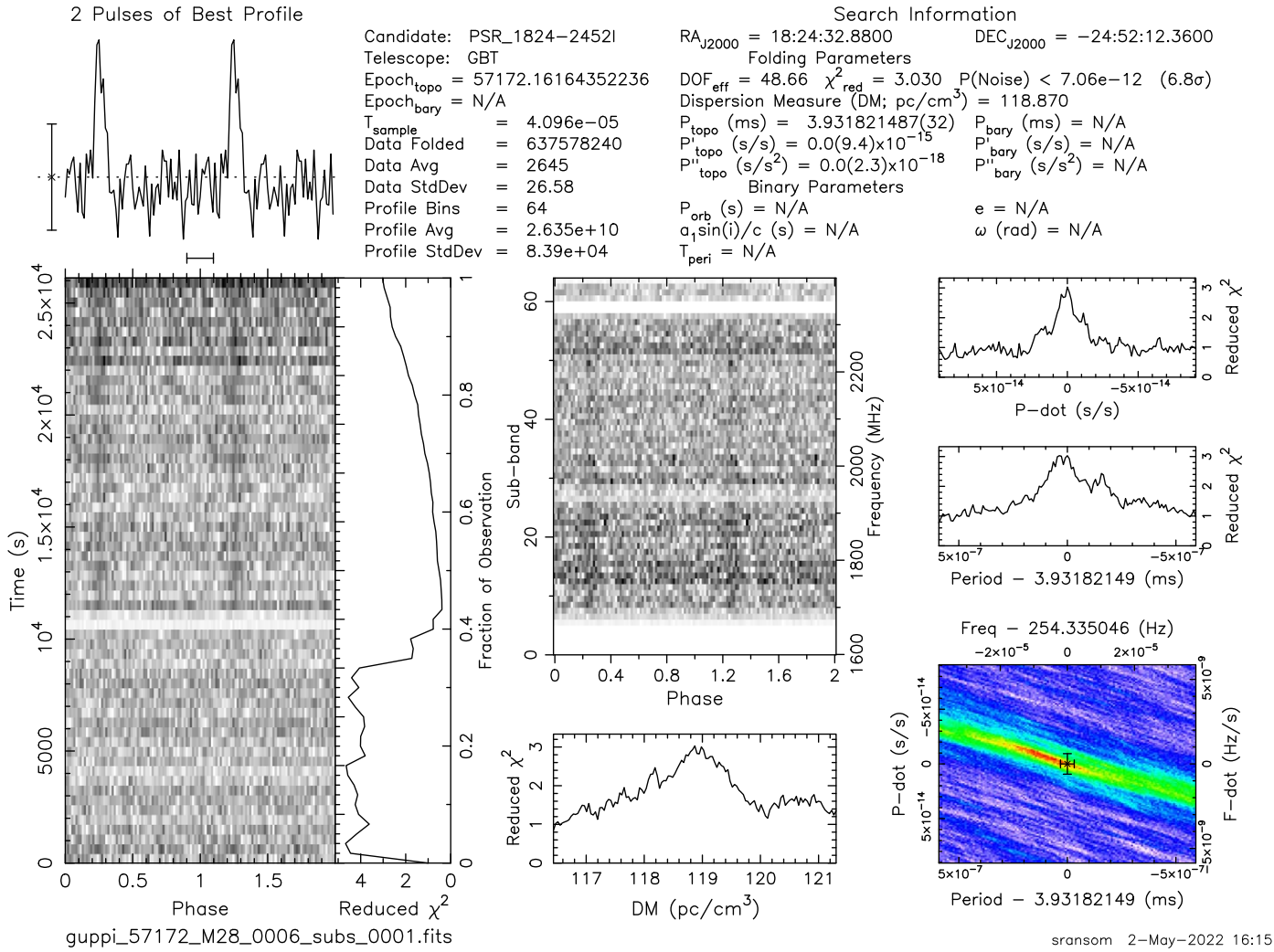

**Figure 7.** GBT plus GUPPI detection of M28I on 57172, during one of the ∼8 hr duration observations of M28, simultaneous with Chandra X-ray observations. The pulsar can be clearly seen coming out of the eclipse in the pulse phase in the bottom left panel vs. time grayscale plot on the left. The integrated pulse profile is shown at the top left. The detection was made using `prepfold` from the `PRESTO` package after optimizing the predicted orbital phasing using `SPIDER_TWISTER`.

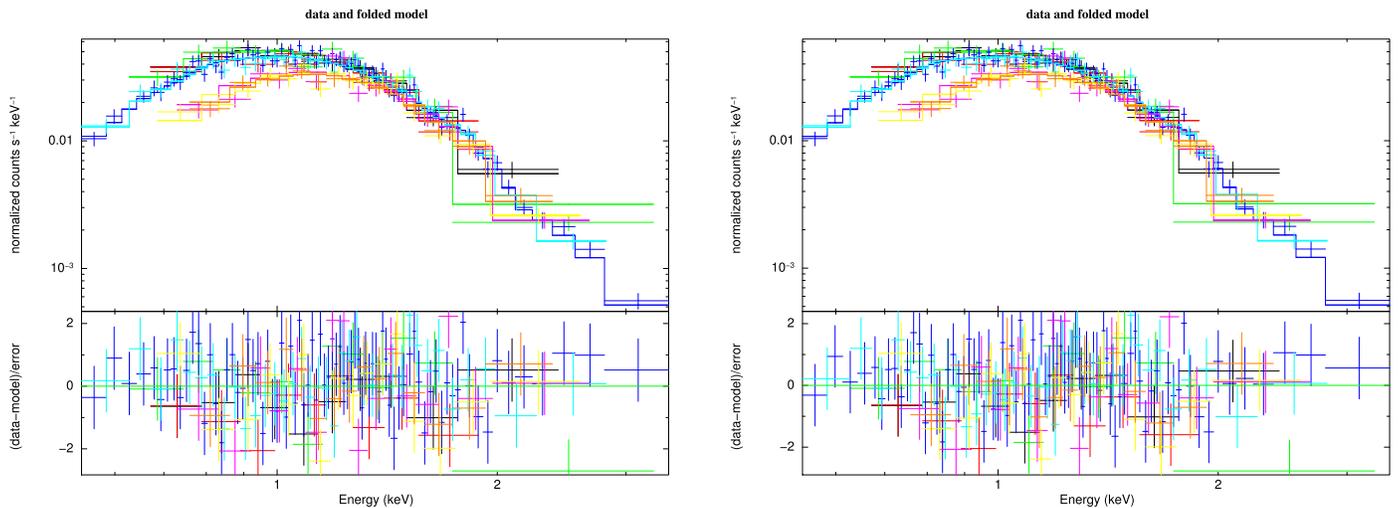

**Figure 8.** X-ray spectra of the qLMXB including all Chandra–ACIS observations. Left panel: fitted to a hydrogen atmosphere model (PILE-UP(TBABS*NSATMOS)). Right panel: fitted using a helium atmosphere model (PILE-UP(TBABS*NSX)).

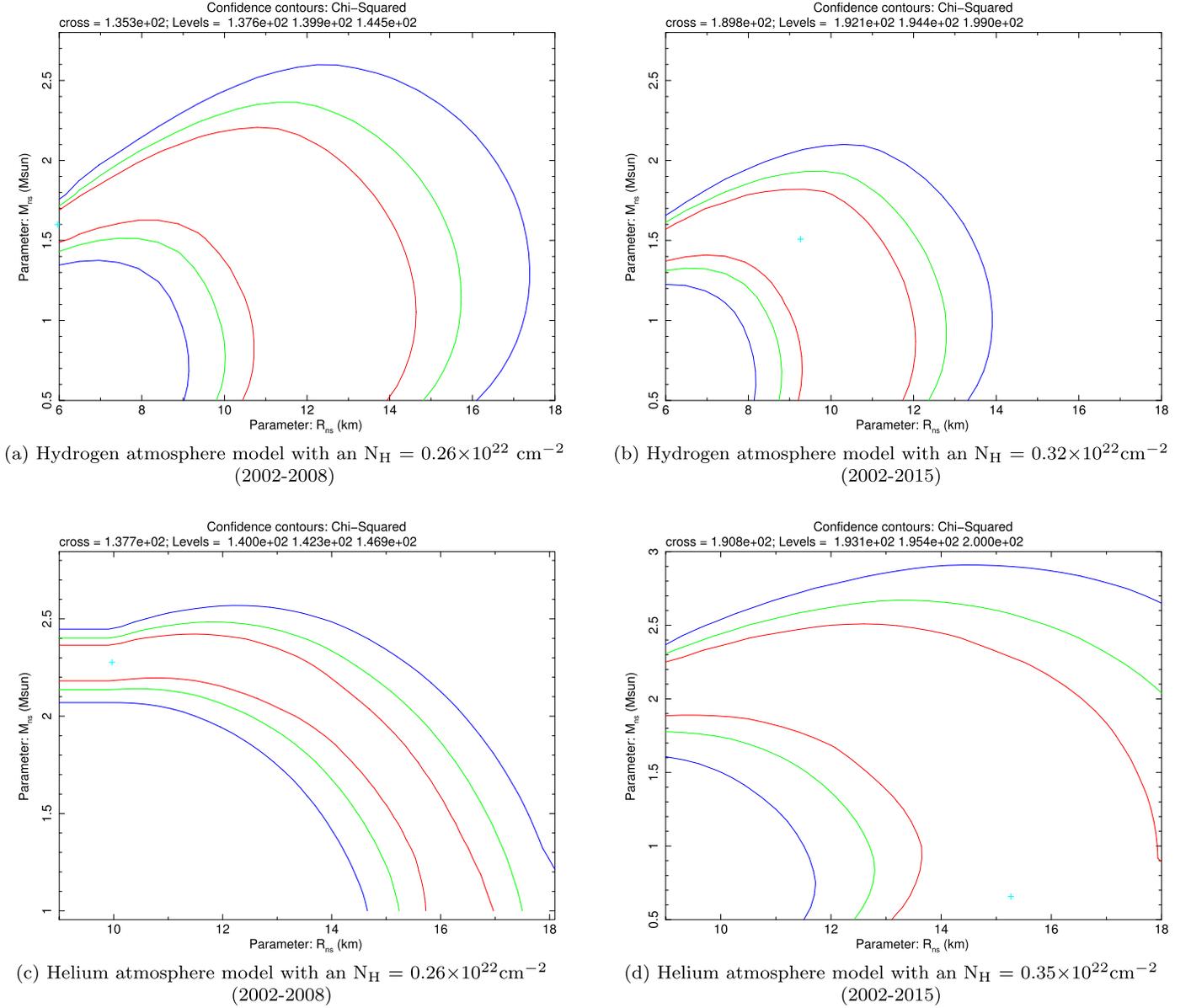

**Figure 9.** Confidence levels (68%, 90%, and 95% in red, green, and blue, respectively) for the mass and radius constraints of the qLMXB, using hydrogen (top panel) and helium (bottom panel) atmosphere models. Left panels: only 2002 and 2008 Chandra observations are included. Right panels: all Chandra observations are used.

Here, $F_{\max}$ and $F_{\min}$ are the maximum and minimum X-ray fluxes, and $EF_{\max}$ and $EF_{\min}$ are their corresponding errors (see, e.g., Saeedi et al. 2022).

We find that 13 of the 46 brightest X-ray sources are variable based on the threshold $S>3$ (marked with "v" in Table 4). Among these, six sources were already identified as variable in previous studies (4, 17, M28L, M28I, 29, and 32; see Becker et al. 2003 and Papitto et al. 2013). Source 21, in a crowded region inside the core of M28, is likely contaminated by the nearby and variable M28L (see Figure 5), so its variability is questionable and flagged with a question mark in Table 4. We thus find six new variable X-ray sources in M28, namely 1, 16, 20, 25, 31, and 33. We show zoomed multiepoch ACIS images of these new variables in Figure 10, together with their best-fit $L_X$ and $\Gamma$. We also mark their locations in Figure 1.

## 4. Discussion

### 4.1. A Compilation of Orbital Variability in Spiders

Compact binary MSPs can shed light on the physics of pulsar winds and relativistic shock acceleration. At the moment of writing, there are 42 known compact binary MSPs within 19 Galactic GCs (Freire 2021) (16 RBs and 26 BWs), and about 50 spider MSPs are known in the Galactic field (Linares & Kachelrieß 2021). The X-ray spectra from these systems can be described by a combination of thermal and nonthermal emission components, originating from the heated polar caps, the NS magnetosphere, and the IBS. Nonthermal X-rays from the accelerated particles in the shock region can be modulated

**Table 4**
Variability of X-Ray Sources in M28

| # | $F_{max}-F_{min}$ | $(EF_{max}^2+EF_{min}^2)^{1/2}$ | $S$[a] | Var[b] |
|---|---|---|---|---|
| 1 | 3.10 | 0.90 | 3.2 | v[c] |
| 2 | 1.50 | 1.30 | 1.1 | |
| 3 | 0.23 | 0.14 | 1.6 | |
| 4 | 24.0 | 5.0 | 5.3 | v |
| 5 | 0.20 | 0.09 | 2.3 | |
| 6 | 0.32 | 0.25 | 1.3 | |
| 7 | 0.13 | 0.14 | 0.9 | |
| 8 | 0.38 | 0.14 | 2.7 | |
| 9 | 0.45 | 0.19 | 2.4 | |
| 10 | 3.22 | 0.37 | 1.9 | |
| 11 | 0.29 | 0.17 | 1.7 | |
| 12 | 1.10 | 0.40 | 2.5 | |
| 13 | 0.22 | 0.09 | 2.5 | |
| 14 | 0.78 | 0.34 | 2.3 | |
| 15 | 0.50 | 0.35 | 1.4 | |
| 16 | 0.85 | 0.26 | 3.3 | v[c] |
| 17 | 6.40 | 1.40 | 4.5 | v |
| 18 | 0.85 | 1.55 | 0.5 | |
| 19 | 9.52 | 0.23 | 2.0 | |
| 20 | 2.30 | 0.70 | 3.4 | v[c] |
| 21 | 2.90 | 0.40 | 6.4 | v? |
| 22 | 6.40 | 1.40 | 4.6 | v |
| 23 | 48.0 | 10.0 | 44.6 | v |
| 24 | 0.73 | 0.29 | 2.5 | |
| 25 | 3.50 | 0.50 | 7.2 | v[c] |
| 26 | 3.25 | 0.80 | 2.7 | |
| 27 | 0.11 | 0.30 | 0.3 | |
| 28 | 4.50 | 1.90 | 2.3 | |
| 29 | 4.80 | 0.60 | 8.4 | v |
| 30 | 1.00 | 0.50 | 2.2 | |
| 31 | 0.92 | 0.27 | 3.4 | v[c] |
| 32 | 3.30 | 0.90 | 3.5 | v |
| 33 | 0.79 | 0.21 | 3.8 | v[c] |
| 34 | 0.19 | 0.18 | 1.1 | |
| 35 | 0.23 | 0.12 | 1.8 | |
| 36 | 0.21 | 0.16 | 1.3 | |
| 37 | 0.49 | 0.30 | 1.6 | |
| 38 | 0.93 | 0.36 | 2.5 | |
| 39 | 0.14 | 0.10 | 1.3 | |
| 40 | 0.57 | 0.40 | 1.4 | |
| 41 | 0.23 | 0.17 | 1.3 | |
| 42 | 0.91 | 0.78 | 1.2 | |
| 43 | 0.11 | 0.17 | 0.7 | |
| 44 | 0.16 | 0.18 | 0.9 | |
| 45 | 0.22 | 0.17 | 1.2 | |
| 46 | 0.19 | 0.14 | 1.4 | |

**Notes.** The first column shows the source ID number from Becker et al. (2003).
[a] Significance of the flux variations.
[b] Variability.
[c] Newly discovered variable sources. $F_{max}$ and $F_{min}$ indicate maximum and minimum fluxes in units of $10^{-14}$ erg cm$^{-2}$ s$^{-1}$. $EF_{max}$ and $EF_{min}$ are their corresponding errors.

at the orbital period (e.g., Huang et al. 2012; Hui et al. 2015). This X-ray orbital modulation found in some BWs and RBs, attributed to the emission from the IBS, has been generally explained by a combination of the Doppler boosting of the flow within the shock, synchrotron beaming, and obscuration by the companion (Bogdanov et al. 2005, 2011).

More recently, Wadiasingh et al. (2017) pointed out a dichotomy in the orbital-phase centering of the DP maximum: most spiders (where orbital X-ray modulation has been measured) have this maximum flux centered on the pulsar's IC, while a few have it centered around the SC. In their model and interpretation, this is due to the intrabinary shock being curved around the pulsar and companion star, respectively. Doppler-boosted synchrotron emission along the shock intersecting the line of sight produces the two peaks in this scenario (see also Wadiasingh et al. 2018). Knowing the location and geometry of the intrabinary shock is important (among other reasons) to quantify the amount of intercepted and reaccelerated particles (e.g., positrons; Linares & Kachelrieß 2021).

We show in Figure 11 the known spider population in the Galactic field and highlight the systems with detected X-ray orbital modulation (open black squares). Note that X-ray orbital modulation has been reported in the literature from both RB and BW spiders spanning a range of $L_X$ ($1.9 \times 10^{30}$–$1.5 \times 10^{32}$ erg s$^{-1}$) and photon index (0.9–2.9). This suggests that there is no detected orbital X-ray variability when the photon index is larger than 3. This could be due to fainter IBS emission so that the thermal component dominates. Indeed, most spiders with $L_X < 4 \times 10^{30}$ erg s$^{-1}$ have $\Gamma > 3$ (see Figure 11).

### 4.1.1. The Transitional RB M28I

In Figure 2, the orbital X-ray light curve of M28I shows a DP orbital modulation centered around the pulsar's IC ($\phi = 0.75$; when the pulsar is between the companion and the Earth). The maxima of the X-ray modulation are found at orbital phases $\phi = 0.6$ and $\phi = 1.0$. Between the two maxima, there is a dip around $\phi = 0.9$. Thus, based on the models of Wadiasingh et al. (2017), we infer that the intrabinary shock in M28I is curved or "wrapped" around the pulsar. We find a peak separation of 0.4 in the orbital X-ray light curve of M28I, similar to what Archibald et al. (2010) found for the RB PSR J1023+0038. In a Doppler-boosted shock with small opening angle (Wadiasingh et al. 2017), the amplitude of the modulation is positively correlated with the inclination of the orbital plane, with the maximum possible modulation corresponding to an inclination angle of 90° (Cho et al. 2018). We find that the orbital modulation of M28I shows a remarkably high fractional semiamplitude (71%; Section 3.1.1), which may be due to a high (nearly edge-on) inclination.

As a tMSP, M28I also experiences transitions from a rotation-powered to an accretion-powered or "outburst" state, as well as an intermediate subluminous disk state with high/active and low/passive modes (Papitto et al. 2013; Linares et al. 2014b; Papitto & de Martino 2022). Figure 12 shows an overview of its $L_X$ in the different states, as revealed by Chandra observations in 2002, 2008, 2013, and 2015. We find that M28I was in the pulsar state in 2015, and we detect for the first time radio pulsations simultaneously with an X-ray observation. So far, M28I was detected in the disk state in two of the four epochs when it has been observed with Chandra: in 2008 and 2013. Cohn et al. (2013) and Pallanca et al. (2013) found another possible occurrence of the disk state in 2009, using the Hubble Space Telescope (HST) optical observations of M28. At present, Chandra observations are the most reliable and efficient way of constraining the duty cycle of these two states. The detection of radio pulsations with GBT reveals the pulsar state (by definition), but a nondetection does not allow a state identification since the pulsar is occulted/eclipsed for a large fraction of the orbit. This can be seen in the

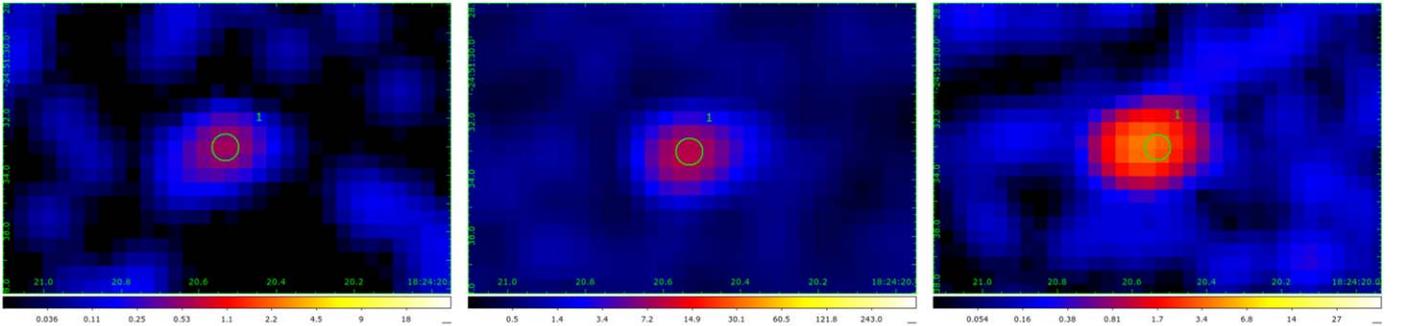
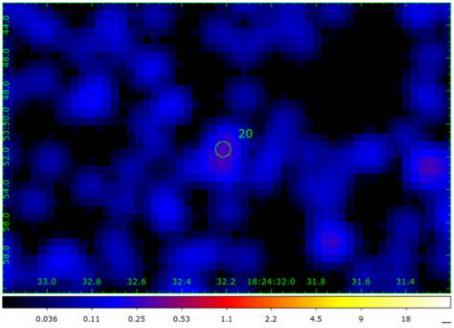 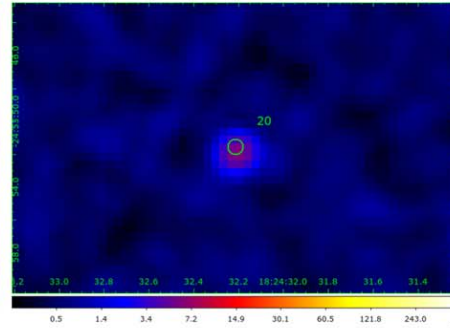 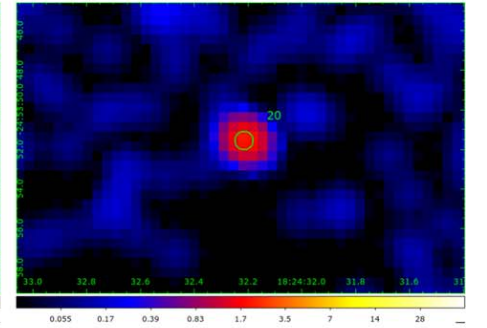

(a) Source 1 (OBS ID:2685).
$L_X = (1.8 \pm 0.8) \times 10^{30}$ erg s$^{-1}$
$\Gamma = 2.3 \pm 0.6$

(b) Source 1 (OBS ID:9132).
$L_X = (4.6 \pm 0.5) \times 10^{31}$ erg s$^{-1}$
$\Gamma = 1.9 \pm 0.1$

(c) Source 1 (OBS ID:16750).
$L_X = (1.1 \pm 0.2) \times 10^{32}$ erg s$^{-1}$
$\Gamma = 1.9 \pm 0.2$

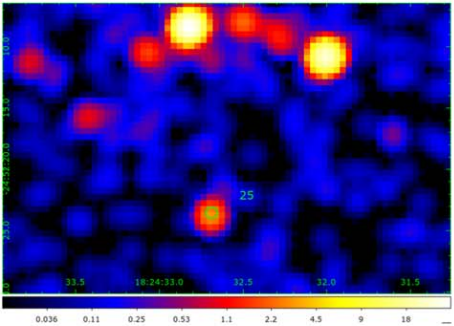 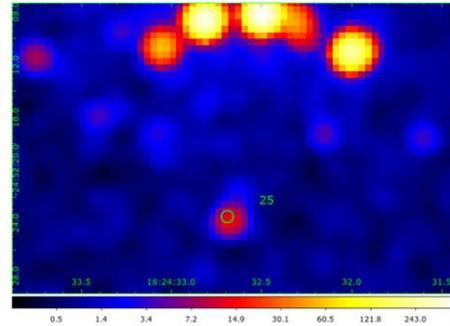 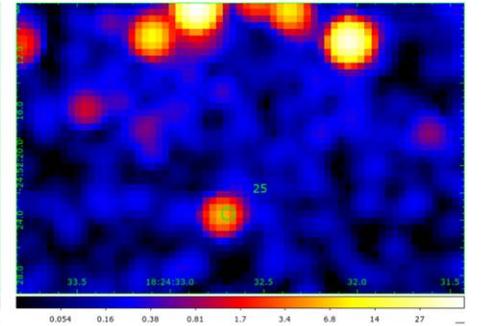

(d) Source 20 (OBS ID:2685).
$L_X = (6.1 \pm 5.0) \times 10^{30}$ erg s$^{-1}$
$\Gamma = 2.5 \pm 1.1$

(e) Source 20 (OBS ID:9132).
$L_X = (3.1 \pm 0.5) \times 10^{31}$ erg s$^{-1}$
$\Gamma = 1.5 \pm 0.3$

(f) Source 20 (OBS ID:16749).
$L_X = (9.1 \pm 0.2) \times 10^{31}$ erg s$^{-1}$
$\Gamma = 1.6 \pm 0.4$

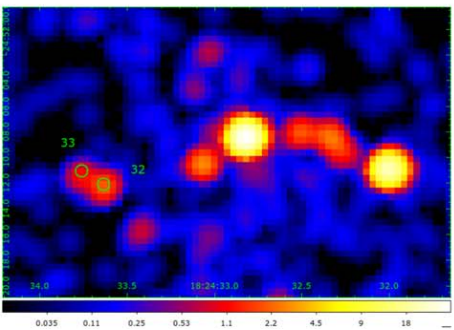 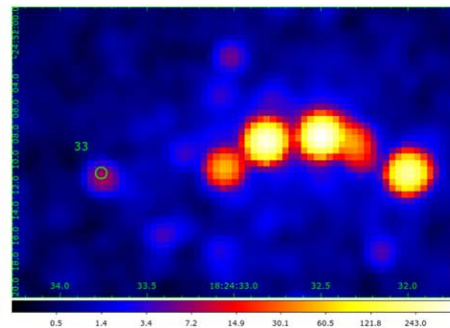 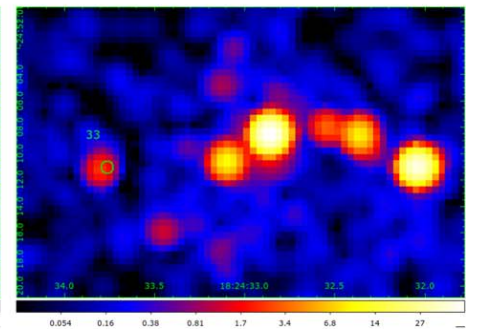

(g) Source 25 (OBS ID:2685).
$L_X = (1.0 \pm 0.3) \times 10^{32}$ erg s$^{-1}$
$\Gamma = 1.2 \pm 0.2$

(h) Source 25 (OBS ID:9132).
$L_X = (2.6 \pm 0.3) \times 10^{31}$ erg s$^{-1}$
$\Gamma = 2.0 \pm 0.1$

(i) Source 25 (OBS ID:16750).
$L_X = (1.1 \pm 0.2) \times 10^{32}$ erg s$^{-1}$
$\Gamma = 1.6 \pm 0.2$

(j) Source 33 (OBS ID:2684).
$L_X = (2.7 \pm 2.0) \times 10^{31}$ erg s$^{-1}$
$\Gamma = 2.1 \pm 0.7$

(k) Source 33 (OBS ID:9132).
$L_X = (7.8 \pm 0.2) \times 10^{30}$ erg s$^{-1}$
$\Gamma = 2.5 \pm 0.2$

(l) Source 33 (OBS ID:16750).
$L_X = (2.9 \pm 0.8) \times 10^{31}$ erg s$^{-1}$
$\Gamma = 2.1 \pm 0.3$

**Figure 10.** Newly discovered variable sources in M28. Chandra–ACIS archival observations of six sources in three epochs (from left to right): 2002, 2008, and 2015. Green circles show the sources from Becker et al. (2003).

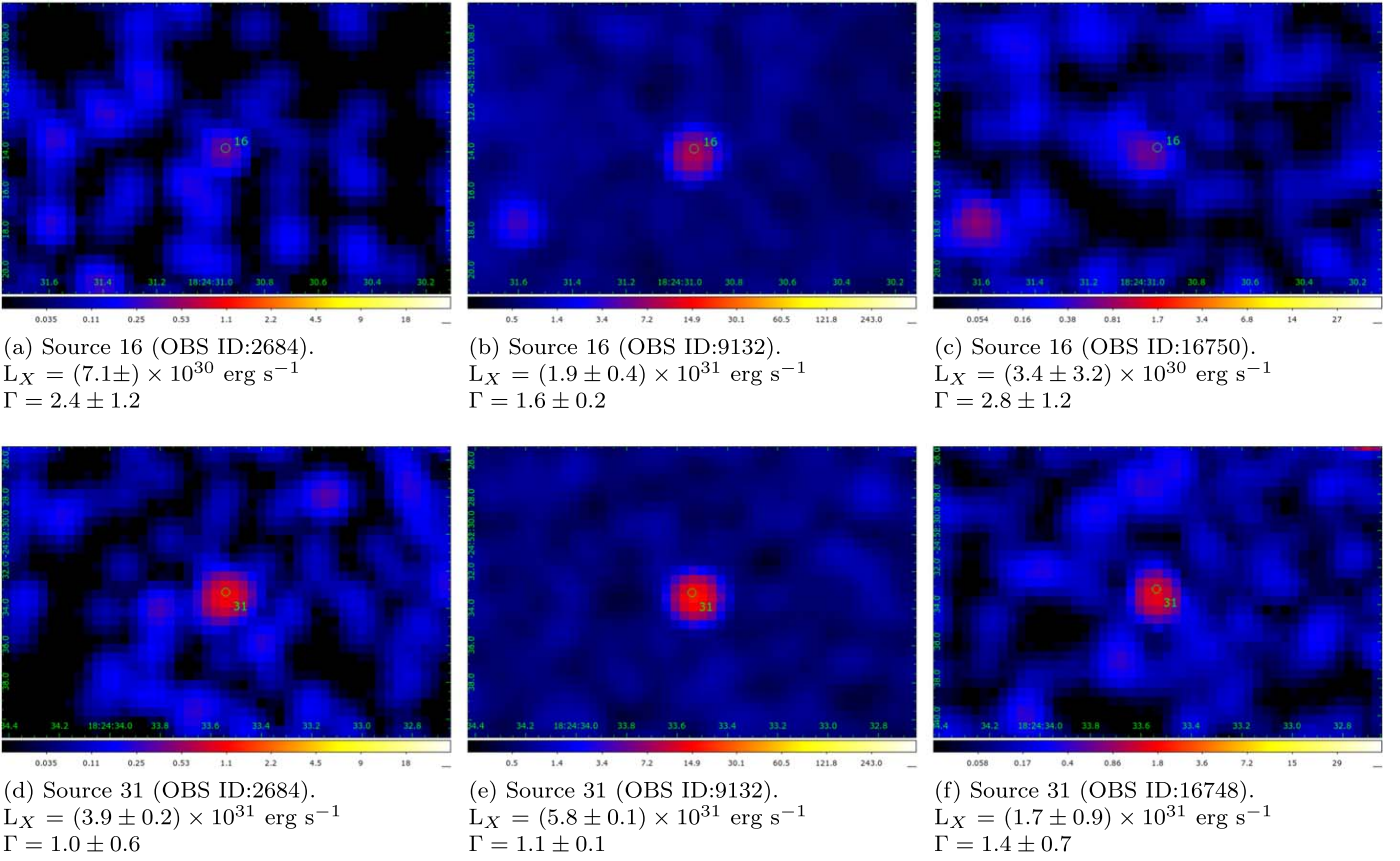

(a) Source 16 (OBS ID:2684).
$L_X = (7.1\pm) \times 10^{30}$ erg s$^{-1}$
$\Gamma = 2.4 \pm 1.2$

(b) Source 16 (OBS ID:9132).
$L_X = (1.9 \pm 0.4) \times 10^{31}$ erg s$^{-1}$
$\Gamma = 1.6 \pm 0.2$

(c) Source 16 (OBS ID:16750).
$L_X = (3.4 \pm 3.2) \times 10^{30}$ erg s$^{-1}$
$\Gamma = 2.8 \pm 1.2$

(d) Source 31 (OBS ID:2684).
$L_X = (3.9 \pm 0.2) \times 10^{31}$ erg s$^{-1}$
$\Gamma = 1.0 \pm 0.6$

(e) Source 31 (OBS ID:9132).
$L_X = (5.8 \pm 0.1) \times 10^{31}$ erg s$^{-1}$
$\Gamma = 1.1 \pm 0.1$

(f) Source 31 (OBS ID:16748).
$L_X = (1.7 \pm 0.9) \times 10^{31}$ erg s$^{-1}$
$\Gamma = 1.4 \pm 0.7$

**Figure 10.** (Continued.)

right panel of Figure 12, where we show an $L_X$ measurement indicative of the pulsar state (Chandra), strictly simultaneous with a radio pulsar nondetection (GBT). As seen in Figure 6, the X-ray photon index $\Gamma$ is approximately constant throughout the pulsar–disk–pulsar state transitions (the fitted spectra are shown in Figure A1, panels (c) and (d)).

*4.1.2. The RB: M28H*

The eclipsing binary pulsar M28H is in a 10.4 hour circular orbit around a nondegenerate star with a minimum inferred mass of $0.17 M_\odot$. The orbital separation of the system is $2.9 R_\odot$ for an assumed inclination of 60° (Bégin 2006). We find an $L_X$ for this RB of $[2.3 \pm 0.4] \times 10^{31}$ erg s$^{-1}$ for a 5.5 kpc distance, consistent with those measured in other RBs and BWs.

In contrast with M28I, we find an orbital modulation with one single peak in the orbital X-ray light curve of M28H. As seen from Figure 3, the maximum occurs when the pulsar is at IC ($\phi = 0.75$), and we find a broad minimum of the X-ray emission around SC ($\phi = 0–0.3$) where the radio eclipses are seen (Bogdanov et al. 2011).

*4.1.3. The BWs: M28G, M28J, and M28L*

The BW MSPs M28G and M28J, namely J1824−2452G and J1824−2452J, are in the core of the GC with periods of 5.9 ms and 4.0 ms, respectively. They have companions with very low masses of $0.011 M_\odot$ and $0.015 M_\odot$ for M28G and M28J, respectively. We do not detect orbital variability in the X-ray light curves of these two BWs (Figure 4). We did not find a reliable orbital modulation for the BW MSP M28L. This may be due to contamination from nearby sources (Figure 5).

*4.2. Neutron Star Mass and Radius*

Our analysis showed that the qLMXB is in a long quiescent regime, and its luminosity remains stable over the 13 yr. For both hydrogen and helium models, we obtained fit parameters that are consistent with the expected value range for a typical NS (see Figure 9). For the hydrogen atmosphere model and $M = 1.4$ $M_\odot$, our constraint on the radius is in the range $R = 9.2–11.5$ km at 68% confidence level. From the helium model, we finder higher radii for a 1.4 $M_\odot$ NS in the range $R = 13.0–17.5$ km. We note that our updated He model constraints are broader and consistent with lower values of $R$ when including the full updated data set, compared to the previous constraints obtained from the 2002–2008 data (Servillat et al. 2012). We conclude that, as noted by Ho & Heinke (2009) and Servillat et al. (2012), the composition of the NS atmosphere is still the main systematic uncertainty in determining $M$ and $R$. Other systematic effects, which we do not explore in this work, include the presence of hot spots, distance uncertainty, abundances of the interstellar medium, and absolute flux calibration (Heinke et al. 2014; Bogdanov et al. 2016; Steiner et al. 2018). We also found different temperatures in the hydrogen and helium atmosphere models, $0.13 \pm 0.01$ keV and $0.10 \pm 0.01$ keV, respectively (at $1\sigma$ confidence level).

Recently, joint NICER and XMM-Newton Observatory measurements have given constraints on $R$ with a different method: pulse-profile modeling with rotating hot spot models (Miller et al. 2021; Riley et al. 2021). Their reported radius $R = 13.7^{+2.6}_{-1.5}$ km is formally consistent with both our H and He constraints. An independent measurement of the NS atmospheric composition would improve the constraints on $M$ and $R$ from this and other thermally emitting qLMXBs.

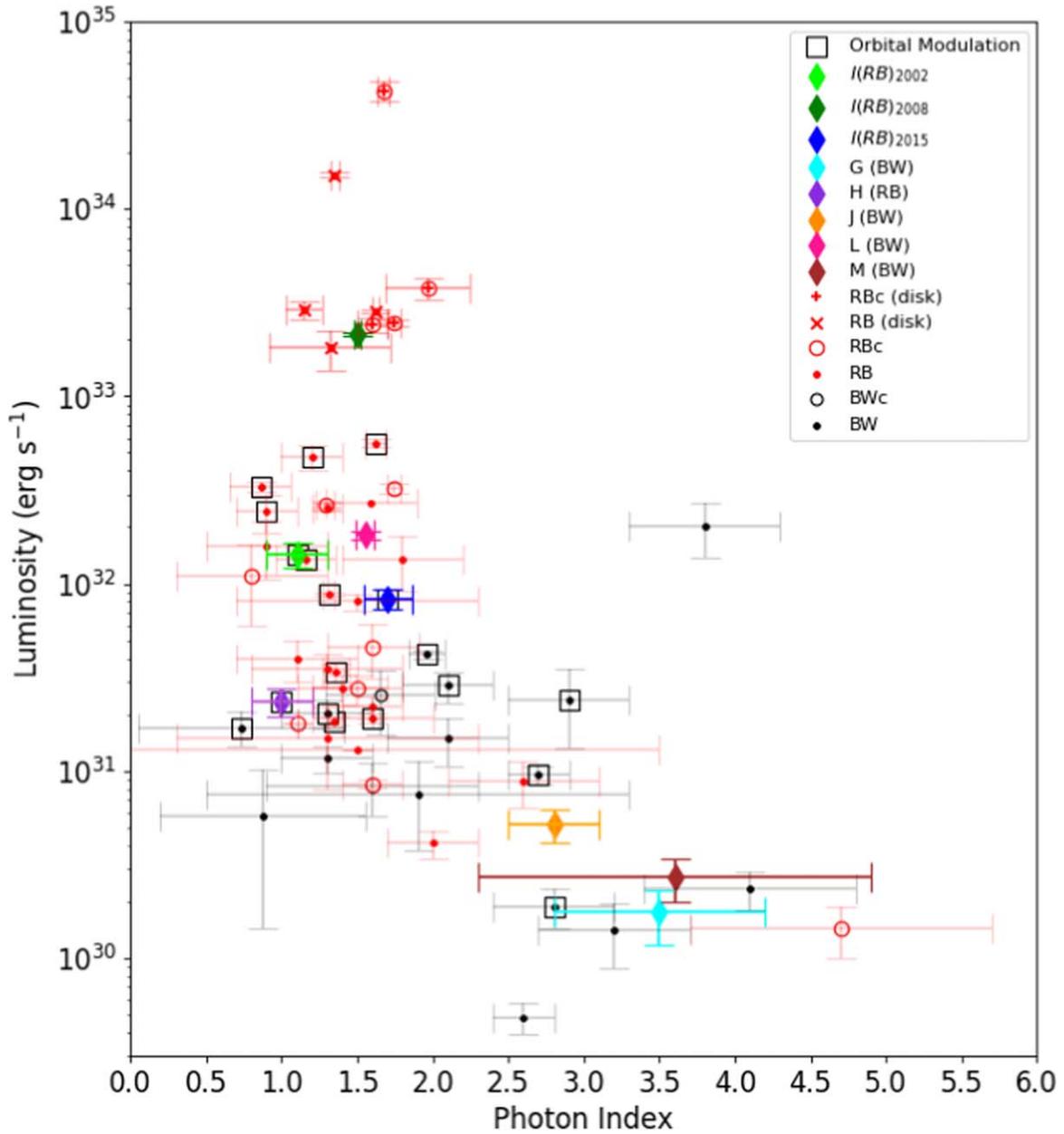

**Figure 11.** Photon index vs. X-ray luminosity ($L_X$) including BWs (filled black circles) and BW candidates (BWc; open black circles), RBs (filled red circles) and RB candidates (RBc; open red circles) in literature together with the six spiders (diamond symbols) analyzed in this work. Red cross symbols and plus symbols show disk states for the RBs and RB candidates, respectively. We include data from Bogdanov et al. (2021). Open squares indicate the systems with orbital modulation detected in the X-ray: J1824−2452H (Bogdanov et al. 2011); J2129−0429 (Hui et al. 2015); J1023 + 0038 (Archibald et al. 2010; Tam et al. 2010); J1227−4853 (de Martino et al. 2020); J1723−2837 (Hui et al. 2014); J1306−40 (Linares 2017); J2039−5618 (Salvetti et al. 2015); J1628−3205 (Roberts et al. 2015); J1311−3430, J1446 −4701 (Arumugasamy et al. 2015); B1957 + 20 (Huang et al. 2012); J2241−5236 (An et al. 2018); J1124−3653, J2256−1024 (Gentile et al. 2014); J1748−2446P, J1748−2446ad, J1748−2446O (Bogdanov et al. 2021); J1824−2452I (this work).

### 4.3. New Faint and Variable X-Ray Sources

GCs are rich environments in terms of interacting binary systems such as LMXBs, cataclysmic variables (CVs), active binaries (ABs), MSPs, and perhaps black hole binaries (Verbunt & Lewin 2005; Bahramian et al. 2020). In the search for variable sources in the GC M28 through the years 2002–2015, we have found six new variable sources with luminosities $L_X < 10^{33}$ erg s$^{-1}$ (see Section 3.5 and Figure 10). In this section, we discuss their possible nature. In Figure 13, we show minimum and maximum X-ray luminosity values with photon indices for the variable sources that we detected. We do not see any photon index $\Gamma > 3$, which suggests that the variability is caused by nonthermal emission in these sources. The X-ray luminosities are between $10^{30}$ and $10^{33}$ erg s$^{-1}$. For all these variable sources, as the $L_X$ increases, their photon index and $\Gamma$ increases (except source 32 where $\Gamma$ is constant).

In the case of source 1, which is located outside the half-light radius of the cluster (Figure 1), $L_X$ increases monotonically from $[1.8 \pm 0.8] \times 10^{30}$ to $[1.1 \pm 0.2] \times 10^{32}$ erg s$^{-1}$, i.e., a factor of about 60 over the course of 13 yr. Source 20 shows a similar monotonic increase in $L_X$, by about a factor 30. In both cases the photon index stays approximately constant (within the errors) in the range 1–2.5. This strong variability and relatively high $L_X$ (reaching $10^{32}$ erg s$^{-1}$) is perhaps reminiscent of qLMXBs. While qLMXBs can be strongly variable and reach high $Lx$ values, typically they get harder as they get brighter (see Rutledge et al. 2002; Fridriksson et al. 2010;

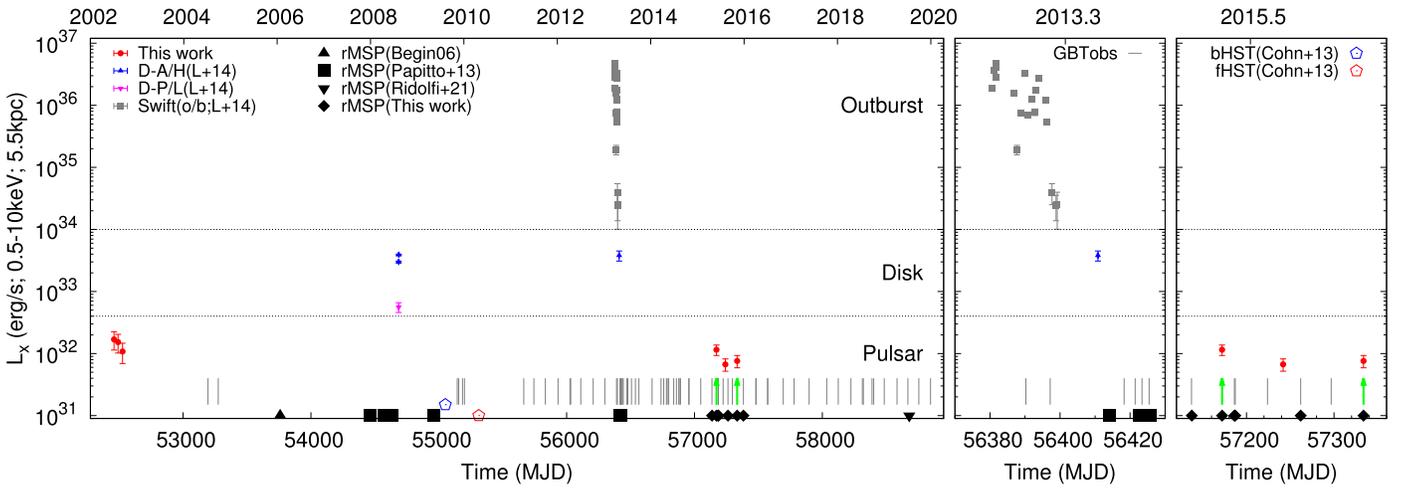

**Figure 12.** Left panel: X-ray luminosity ($L_X$) of the transitional MSP M28I as measured by Chandra and Swift between 2002 and 2015, together with the dates when the radio MSP was detected (shown with filled black symbols labeled "rMSP," with different origins as indicated; vertical gray lines show GBT observation dates). Red circles, blue/magenta triangles, and gray squares show the $L_X$ measurements taken in the pulsar, disk, and outburst states, respectively. The horizontal dashed lines show the approximate $Lx$ boundaries between these states (Linares 2014). Open blue/red pentagons show the bright/faint HST detections indicating that M28I was in the disk/pulsar state in 2009/2010 (Cohn et al. 2013). Middle panel: zoom into the outburst light curve (Swift observations shown with gray squares) showing the detection of a disk-active/high state 10 days after the end of the outburst (blue triangle, from a Chandra - HRC observation; Linares et al. 2014b) and the detection of the rMSP 4 days thereafter (Papitto et al. 2013). Right panel: zoom into the 2015 coordinated (Chandra+GBT) campaign. The green arrows show the strictly simultaneous radio/X-ray observations reported in this work.

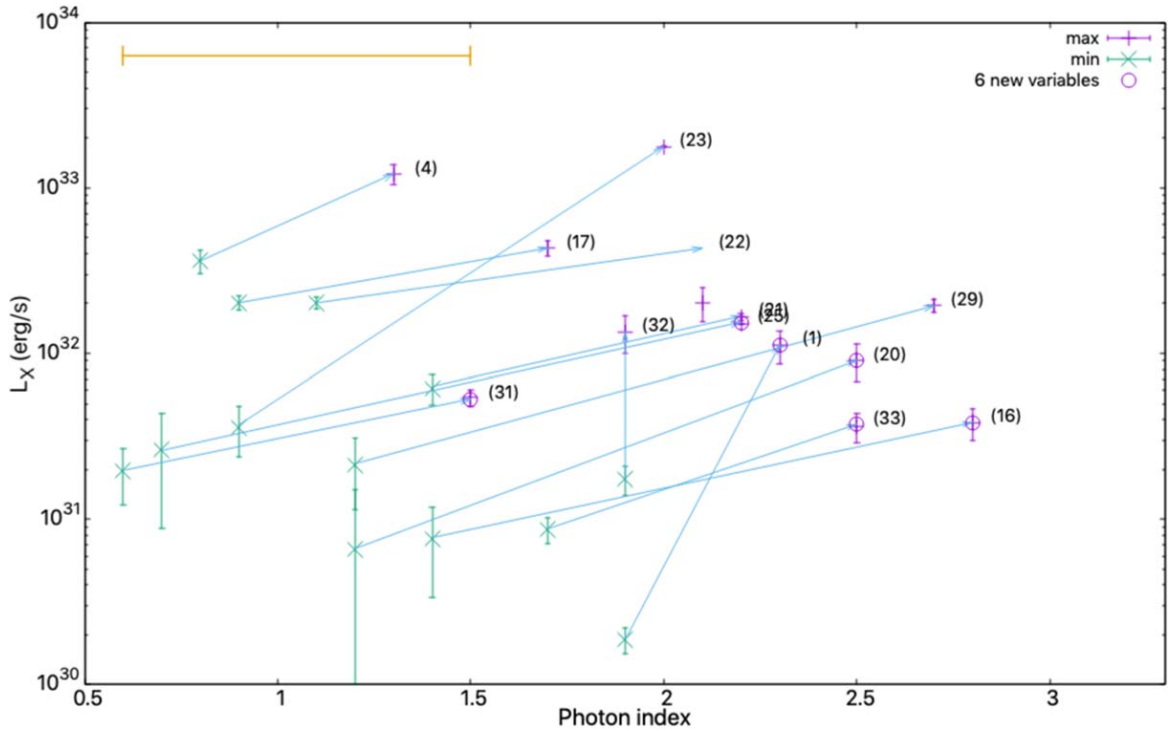

**Figure 13.** Maximum (magenta plus) and minimum (green cross) X-ray luminosities vs. photon index for the 13 detected variable sources in M28. Source ID numbers are shown between parentheses, and the six newly discovered variables are marked with red circles. The yellow line at the top left corner represents the average error for the photon index, which corresponds to $\Gamma = \pm 0.45$.

Bahramian et al. 2014; and most generically, Wijnands et al. 2015). Sources 16 and 31 are both just outside the core and increasing their luminosity in 2008, then become fainter again in 2015. Source 25 is within the core radius, and its luminosity fluctuates between $L_X \sim 10^{31}$–$10^{32}$ erg s$^{-1}$, changing by about a factor 4.

In the case of source 33, within the core radius, it is found to be blended with source 32 in the 2002 observation as seen in Figure 10. Source 32 was also detected as variable in 2003 by Becker et al. ([2003](#)), and it is not detected in the 2008 and 2015 observations. Taking advantage of the absence of source 32, we find variability in the flux of source 33 increasing its brightness from 2008 to 2015. In the 2008 epoch, the $L_X$ of source 33 is $[7.8 \pm 0.2] \times 10^{30}$ erg s$^{-1}$ with $\Gamma = [2.5 \pm 0.2]$. In the 2015 epoch, $L_X$ is $[2.9 \pm 0.8] \times 10^{31}$ erg s$^{-1}$ with $\Gamma = [2.1 \pm 0.3]$.

Identifying components at other wavelengths may reveal the true nature of these intriguing variable low-$L_X$ sources. Some CVs are expected among our variable sources since there are from 100 to 1000 times more white dwarfs than NSs in a GC (Maccarone & Knigge [2007](#)). However, since background active galactic nuclei can produce high X-ray/optical flux ratios, they can act as CVs (Bassa et al. [2005](#)). From the observational point of view, an alternative approach to identifying X-ray sources could be to simultaneously combine the data taken in different energy bands (X-ray, UV, optical, IR). In particular, JWST IR and/or HST optical observations may help identify the six newly identified X-ray sources in our Chandra study.

We thank Eric Miller for useful discussion of the ACIS response evolution. We thank the referee for their careful reading and valuable comments that helped to improve our manuscript. E.V. is grateful to the staff of Chandra X-ray Center for their quick responses on the interactive analysis. E.V. warmly thanks Sarp Akcay for his endless support. E.V. acknowledges support from the Spanish MINECO grant AYA2017-86274-P and from the Spanish MICINN/AEI grant PID2020-117252GB-I00. M.L. acknowledges funding from the European Research Council (ERC) under the European Union's Horizon 2020 research and innovation program (grant agreement No. 101002352). S.B. acknowledges support provided by NASA through Chandra Award Number GO5-16050B issued by the Chandra X-ray Center, which is operated by the Smithsonian Astrophysical Observatory for and on behalf of NASA under contract NAS8-03060. A.P. acknowledges financial support from the Italian Space Agency (ASI) and National Institute for Astrophysics (INAF) under agreements ASI-INAF I/037/12/0 and ASI-INAFn.2017-14-H.0, from INAF "Sostegno alla ricerca scientifica main streams dell'INAF", PresidentialDecree 43/2018, and from the Italian Ministry of University and Research (MUR), PRIN 2020 (prot.2020BRP57Z). Pulsar research at UBC is supported by an NSERC Discovery Grant and by the Canadian Institute for Advanced Research. The National Radio Astronomy Observatory is a facility of the National ScienceFoundation operated under cooperative agreement by Associated Universities, Inc. The Green Bank Observatory is a facility of the National Science Foundation operated under cooperative agreement by Associated Universities, Inc. SMR is a CIFAR Fellow and is supported by the NSF Physics Frontiers20Center awards 1430284 and 2020265. In this article, we used the data and software provided by the High Energy Astrophysics Science Archive Research Center (HEASARC). We acknowledge extensive use of NASA's Astrophysics Data System (ADS) Bibliographic Services and the ArXiv. The scientific results reported in this research are based on the observations made by the Chandra X-ray Observatory and data obtained from the Chandra Data Archive.

*Facilities:* Chandra ACIS. GBT.

*Software:* CIAO (Fruscione et al. [2006](#)), HEAsoft (HEASARC [2014](#)), PRESTO.

## Appendix
## Details of the X-Ray and Radio Analysis

In this appendix we present the radio and X-ray positions of the known pulsars, together with their uncertainties and angular separations between the X-ray and radio positions (see Table [A1](#)). We include the individual X-ray spectral fits for the pulsars in M28 (Figures [A1](#), [A2](#), and [A3](#)). We also present the measurements and upper limits on the X-ray luminosity of the faint sources in M28 (Figure [A4](#)). Finally, we show the molecular contamination effect on the long-term count rate light curve of the qLMXB (Figure [A5](#)).

**Table A1**
Radio and X-Ray Positions of the Pulsars in M28 and Their Positional Uncertainty

| Source | R. A.$_{\text{Radio}}$ J2000 | Decl.$_{\text{Radio}}$ J2000 | R. A.$_{\text{X–ray}}$ J2000 | Decl.$_{\text{X–ray}}$ J2000 | $P_{\text{err}}$[a] (arcsec) | $\theta$[b] (arcsec) |
|---|---|---|---|---|---|---|
| A | 18 24 32.00799483(72) | −24 52 10.8348902(28) | 18 24 32.10 | −24 52 10.81 | 0.3 | 0.03 |
| B | 18 24 32.54585781(35) | −24 52 04.3560436(06) | ⋯ | ⋯ | ⋯ | − |
| C | 18 24 32.19250199(14) | −24 52 14.6818430(26) | 18 24 32.20 | −24 52 14.80 | 0.1 | 0.1 |
| D | 18 24 32.42200854(39) | −24 52 26.2224825(13) | 18 24 32.43 | −24 52 26.90 | 0.1 | 0.6 |
| E | 18 24 33.08952070(88) | −24 52 13.4701099(81) | 18 24 33.05 | −24 52 13.30 | 0.1 | 0.6 |
| F | 18 24 31.81278784(22) | −24 49 24.9511809(85) | 18 24 31.80 | −24 49 24.89 | 0.4 | 0.2 |
| G | 18 24 33.02548892(97) | −24 52 17.1927818(36) | 18 24 33.03 | −24 52 17.00 | 0.1 | 0.2 |
| H | 18 24 31.61052125(72) | −24 52 17.2268378(32) | 18 24 31.61 | −24 52 17.35 | 0.3 | 0.1 |
| I | 18 24 32.50368185(81) | −24 52 07.4353327(34) | 18 24 32.51 | −24 52 07.66 | 0.3 | 0.3 |
| J | 18 24 32.73414004(26) | −24 52 10.3208653(08) | 18 24 32.71 | −24 52 10.18 | 0.3 | 0.4 |
| K | 18 24 32.49746490(59) | −24 52 11.3661979(78) | 18 24 32.49 | −24 52 11.31 | 0.3 | 0.1 |
| L | 18 24 32.35856942(90) | −24 52 08.1973300(70) | 18 24 32.34 | −24 52 08.02 | 0.3 | 0.3 |
| M | 18 24 33.1835(5) | −24 52 08.179(23) | 18 24 33.21 | −24 52 08.20 | 0.1 | 0.3 |
| N | 18 24 33.1418(12) | −24 52 11.89(3) | ⋯ | ⋯ | ⋯ | − |

**Notes.**
[a] Positional uncertainty radius.
[b] Angular separation between the X-ray and radio positions. Units of R.A. are hours, minutes, and seconds, and units of decl. are degrees, arcminutes, and arcseconds. M's and N's radio positions are taken from Douglas et al. ([2022](#)). X-ray positions of I, L, and their positional uncertainties (68% c. l.) are taken from Becker et al. ([2003](#)). X-ray positions of C, D, E, G, M, and their positional uncertainties (68% c. l.) are taken from Cheng et al. ([2020](#)). The rest of the X-ray positions and their positional errors are obtained in this work (95% c. l.).

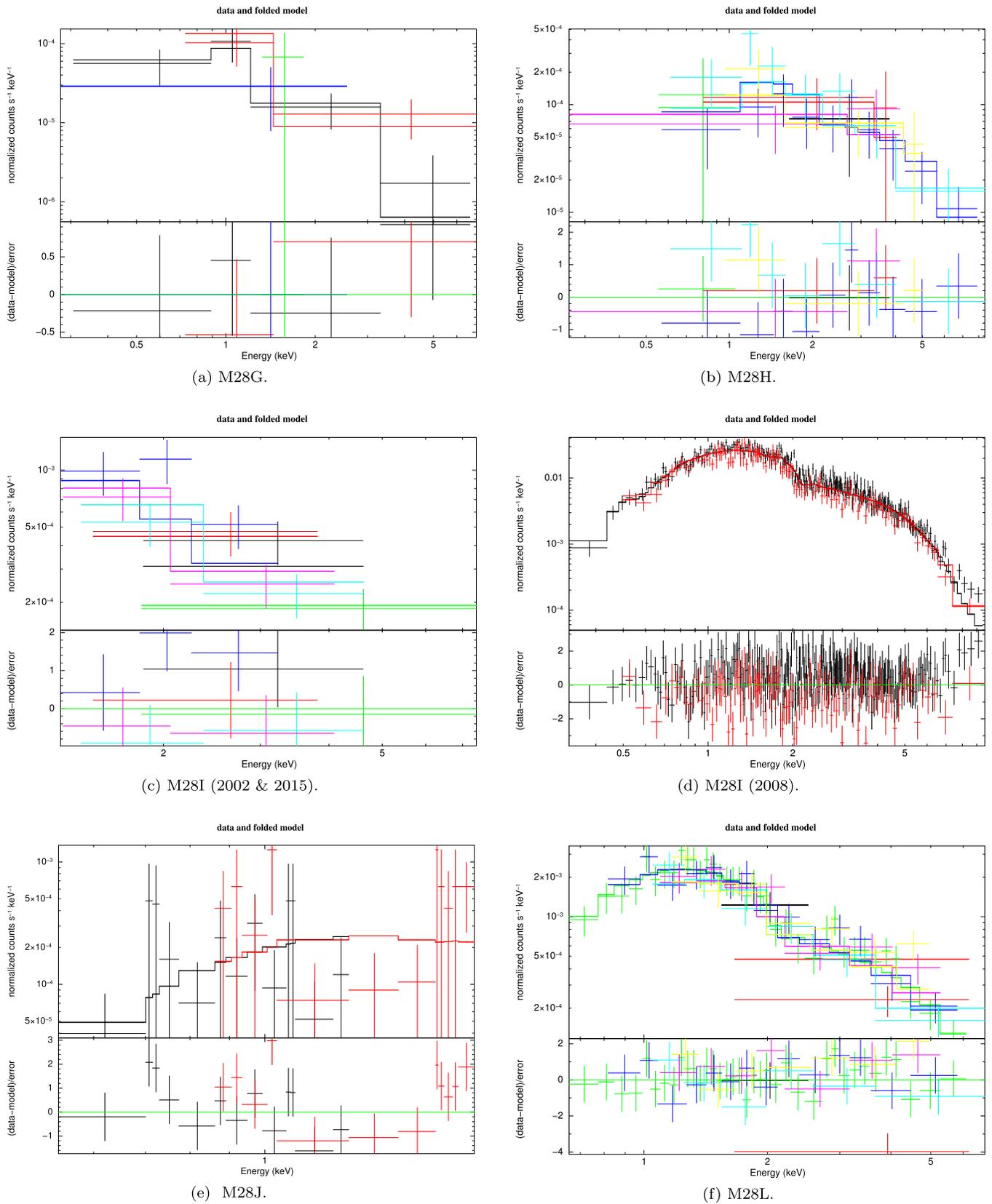

**Figure A1.** Upper panels: X-ray spectra of the spiders in M28. Lower panels: the best-fit residuals.

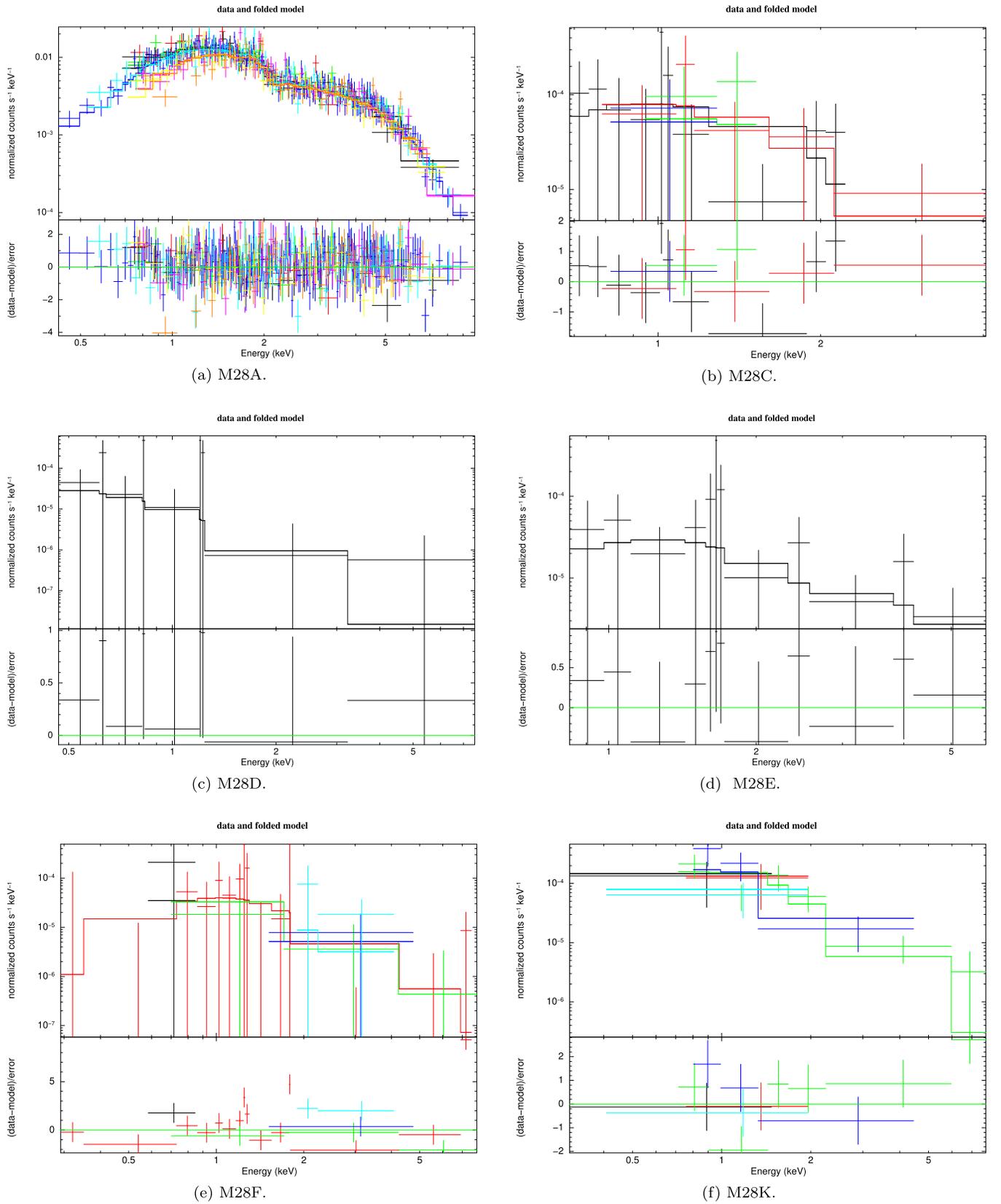

**Figure A2.** Upper panels: X-ray spectra of the rest of the detected pulsars in M28. Lower panels: the best-fit residuals.

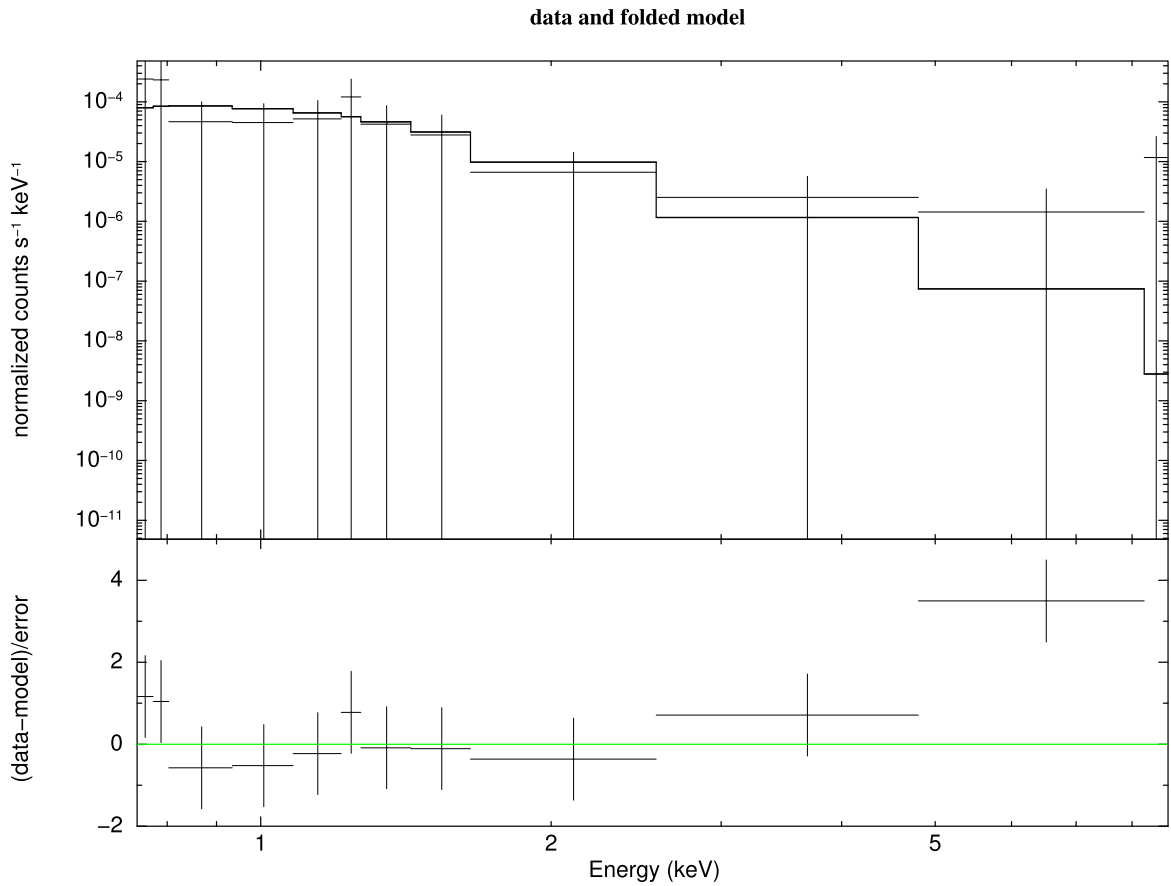

(a) M28M.

**Figure A3.** Upper panel: X-ray spectrum. Lower panel: the best-fit residuals.

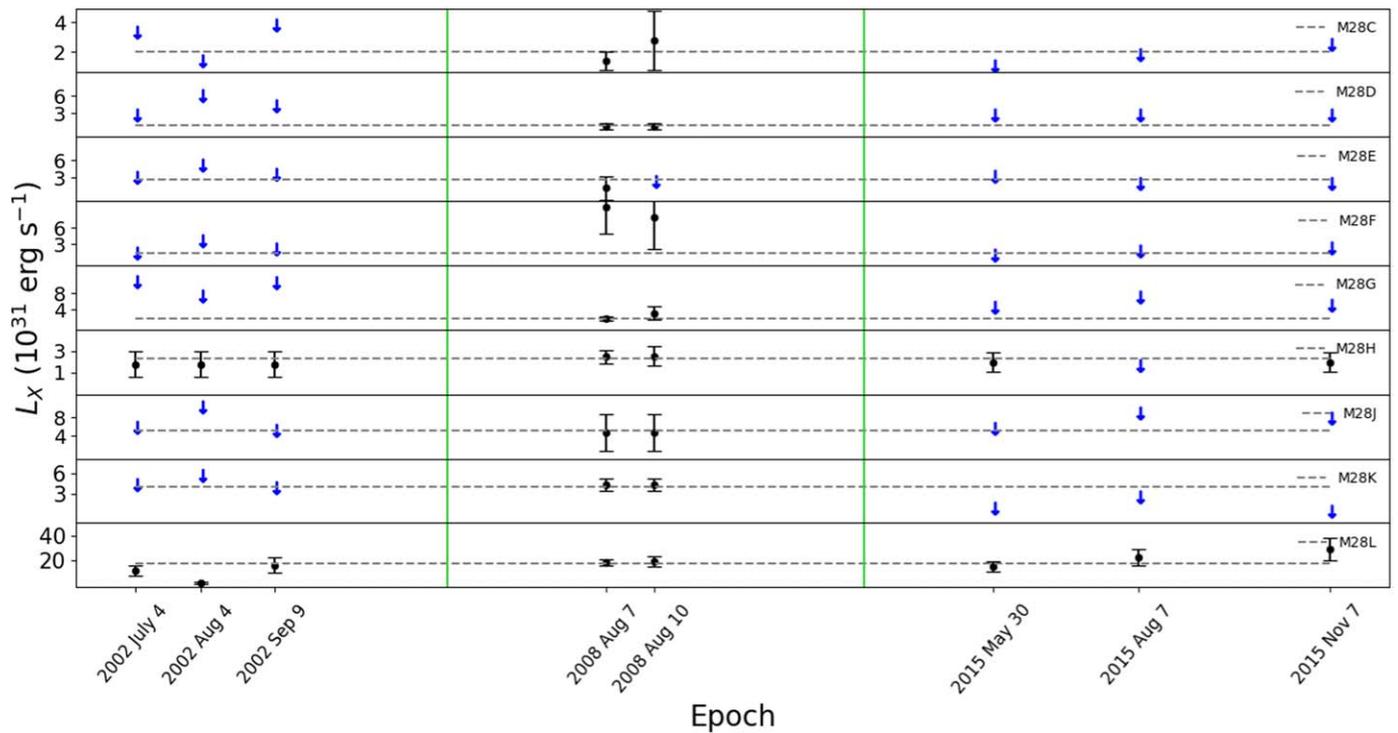

**Figure A4.** Luminosity evolution for the faint pulsars. Blue arrows show upper limits; filled black circles indicate detections. Horizontal dashed lines indicate the average luminosity for individual sources. The green vertical lines separate the epochs visually.

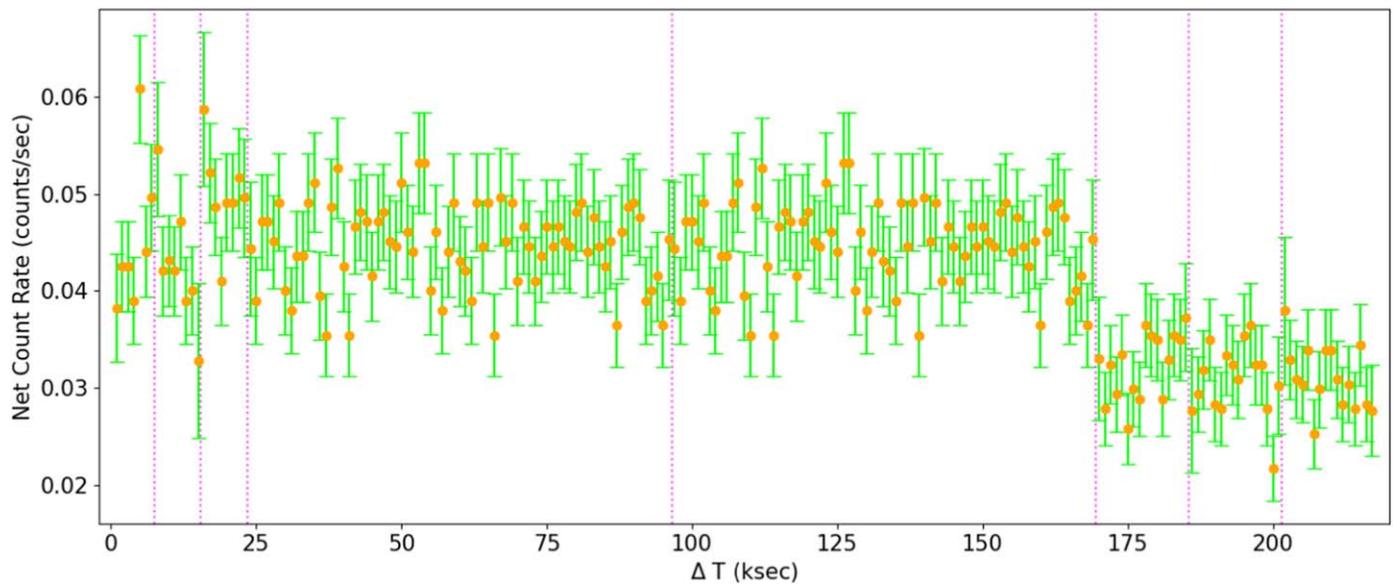

**Figure A5.** X-ray count rate light curve (0.2–10.0 keV) of the qLMXB in M28, including all eight Chandra–ACIS observations (separated by vertical lines, with arbitrary time offsets for display purposes). The molecular contamination effect is apparent as a drop in count rate around $\Delta T \sim 170$ ks.

## ORCID iDs

Eda Vurgun ⓘ https://orcid.org/0000-0001-6544-2713
Manuel Linares ⓘ https://orcid.org/0000-0002-0237-1636
Scott Ransom ⓘ https://orcid.org/0000-0001-5799-9714
Alessandro Papitto ⓘ https://orcid.org/0000-0001-6289-7413
Slavko Bogdanov ⓘ https://orcid.org/0000-0002-9870-2742
Enrico Bozzo ⓘ https://orcid.org/0000-0002-7504-7423
Nanda Rea ⓘ https://orcid.org/0000-0003-2177-6388
Domingo García-Senz ⓘ https://orcid.org/0000-0001-5197-7100
Paulo Freire ⓘ https://orcid.org/0000-0003-1307-9435
Ingrid Stairs ⓘ https://orcid.org/0000-0001-9784-8670